\documentclass[12pt]{article} 
\usepackage{amssymb,amsmath,amsfonts,theorem, subfigure} 
\usepackage{graphicx}
\usepackage{epic,curves}
\usepackage{multirow}

\newcommand{\tri}{\text{tri}}
\newcommand{\hex}{\text{hex}} 
\newcommand{\vac}{|0\rangle}

\theorembodyfont{\itshape} 
\theoremheaderfont{\scshape}
\theoremstyle{plain}  
\newtheorem{Prop}{Proposition}
\newtheorem{Lem}[Prop]{Lemma}
\newtheorem{Cor}[Prop]{Corollary}

\title{Behavior of the two-dimensional Ising model\\
at the boundary of a half-infinite cylinder}
\author{
Yvan Saint-Aubin\footnote{D\'epartement de math\'ematiques et de statistique, Universit\'e de Montr\'eal, C.P.\ 6128, succ.\ centre-ville, Montr\'eal, Qu\'ebec, Canada, H3C 3J7; {\ttfamily saint{\char'100}crm.umontreal.ca}},
Louis-Pierre Arguin\footnote{Courant Institute of Mathematical Sciences,
New York University ,
251 Mercer St.,
New York, NY 10012, USA; {\ttfamily arguin{\char'100}math.nyu.edu}} and
Hassan Aurag\footnote{Email: {\ttfamily
aurag{\char'100}cae.com}}
}
\date{\today}

\begin{document} 
\maketitle

%
%
 
\begin{abstract}
The two-dimensional Ising model is studied at the boundary
of a half-infinite cylinder. The three regular lattices (square,
triangular and hexagonal) and the three regimes (sub-, super- and critical)
are discussed.  The probability of having precisely
$2n$ spinflips at the boundary is computed as a function of the
positions $k_i$'s, $i=1,\dots, 2n,$ of the spinflips.
The limit when the mesh goes
to zero is obtained. For the square lattice, 
the probability of having $2n$ spinflips,
independently of their position, is also computed. 
As a byproduct we recover a result of De Coninck showing that the limiting distribution of the number of spinflips is Gaussian.
The results are obtained as consequences of Onsager's solution and are rigorous.
\end{abstract}

%
%


\section{Introduction}\label{sect:intro}

The present paper gives a rigorous description
of the spin configurations seen at the end of a half-infinite
cylinder covered by the Ising model on square, triangular or
hexagonal lattices. Both
the discrete case and its continuous limit are made explicit. 
Section 2, devoted to the square lattice,
describes the boundary behavior by giving, for a fixed
number $2n$ of spinflips, the probability distribution as a
function of the positions of the spinflips. 
The three regimes, subcritical,
critical and supercritical, are discussed. 
It should be stressed that these probabilities 
{\em are not} correlation functions, even though
they can be used to calculate them. 
In Section 3 we compute, for the square lattice, 
the distribution of the random 
variable (number of spinflips) $/$ (number of
sites at the boundary) at the critical temperature.
Section 4 extends the results of Section 1 to the triangular and hexagonal lattices.
The results (and the methods to obtain them) are simple, though 
non-trivial, and the thermodynamical limits are shown to depend only
on the behavior of smooth functions defined on $[-\pi,\pi]$ for the
square lattice and on $[-\pi/2,\pi/2]$ for the other two. Only their
behavior at zero and at the extremities on these intervals play a role.

The calculation presented here
could have been done by several methods and for other orientations of the lattices. We chose the most classical technique, the one based on the transfer matrix; in the present case, this method turns out to be simple and scientists with expertise in neighboring fields will be able to follow the argument with minimal investment. It is impossible to make a short overview of more than sixty years of developments on the Ising model. There are two results however that we emphasize since they are directly related to ours. Abraham \cite{Abraham2} proved that the limit distribution, with a non-trivial scaling, for the magnetization of the Ising model on a square lattice is Gaussian at the edge of a half-infinite cylinder and at criticality.
Along the same line, De Coninck \cite{deConinck} showed that in the limit the joint distribution for the magnetization and the energy (which is a linear function of the number of spinflips) is also Gaussian. The two proofs use the transfer matrix method to directly compute the characteristic function of the variables. We recover here De Coninck's result for the number of spinflips using the distribution in the discrete case computed in Section 2 and a combinatorial lemma found in the Appendix. We obtain explicit expressions for both the mean and the variance of the limiting distribution.

Finding the relative weights of boundary configurations
as a function of the number   and locations of spinflips might seem
only a mildly interesting exercise. A word of explanation is therefore
in order. 
With the invention of the stochastic Loewner equation (SLE)
by Schramm \cite{Schramm} and proofs by Smirnov \cite{Smirnov}
that percolation and the Ising model are conformally invariant 
in the limit when the mesh goes to zero, new rigorous tools have been available
to probe critical phenomena, and with these tools, new observables have
been introduced. Suppose that boundary conditions are imposed at the 
extremity of the half-infinite cylinder as follows. Let
$0\le \theta_1<\theta_2<\theta_3<\theta_4<2\pi$ be four angles. They define four
intervals along the boundary and we suppose that the sign of Ising spins
are constant on each, but alternate from one interval to the next. 
Such boundary conditions force interfaces, that is contours between
constant-sign clusters, to intersect the boundary at the $\theta$'s, 
and only there. Since these interfaces cannot cross, the interface
starting at $\theta_1$ must end at $\theta_2$ or $\theta_4$.
One can therefore ask what is the probability that the interface
starting at $\theta_1$ goes to $\theta_2$. At least five groups
in the last eight years computed this new ``observable'', and they all had
to solve the problem that is the subject of the present paper.
The first two groups, two of the present authors \cite{ArguinYSA},
and  Bauer, Bernard and Kyt\"ol\"a \cite{BauerBK}, used conformal 
field theory (CFT) to do it, even though the latter group 
was actually interested in devising a way to define multiple SLE processes.
The three other groups used purely SLE methods to obtain the result
\cite{SchrammWilson, Dubedat, KozdronLawler}. We were not able
to compute directly from the lattice models the probability just
described. This is why we turned in \cite{ArguinYSA} to CFT to do
it. And this is why the present paper limits itself 
to computing the relative weights as functions of the number and 
locations of spinflips, disregarding how the interfaces join the
spinflips. Despite its limitations, the calculation has several
redeeming features. First, it is done from first principles
and relies on classical methods. Second, contrarily to the 
results obtained from SLE or CFT, it gives an
explicit result for any number of spinflips, not only for four.
Third, it allows for subcritical and supercritical regimes to
be studied. Finally, it provides information on a mesoscopic scale, 
namely on the distribution of the number of spinflips at the boundary. 
We note also that the continuous distribution for the positions of
four spinflips at the boundary (cf.~equation \eqref{eq:prn2}) was an important 
element in the computation just described \cite{ArguinYSA}.

%
%

\section{Behavior at the boundary for the square lattice}\label{sect:cesaro}

\subsection{Notation}

Let $\sigma:\{1,2,\dots, m\}\rightarrow \{+1,-1\}$ 
be a configuration along the 
circular extremity of a half-infinite cylinder covered by the square
lattice. The number $m$ of sites on this circle is
taken to be even.
It is convenient to encapsulate the
information contained in $\sigma$ in the following data: the value
$s\in\{+1,-1\}$ of $\sigma$ at $1$ and the positions $k_i\in\{1,2,\dots, m\},
1\le i\le 2n$, of spinflips. A spinflip in $\sigma$ occurs at $k_i$ if 
$\sigma_{k_i-1}=-\sigma_{k_i}$ with $\sigma_0=\sigma_m$ and $\sigma_1=
\sigma_{m+1}$ by definition. We choose $1\le k_1<k_2< ... <k_{2n}\le m$. The
number $n$ can also be seen as the number of maximally connected
stretches of $+$-spins along the boundary. We identify the functions
$\sigma$ with their data $(s;k_1,k_2, \dots, k_{2n})$.

If $\sigma$
and $\sigma'$ are two configurations on contiguous circles along the cylinder, the
transfer matrix $T:\mathbb{C}^{2^m}\rightarrow \mathbb{C}^{2^m}$ is given by its
matrix elements
$T_{\sigma;\sigma'}=\exp (\nu \sum_{k=1}^m \sigma_k\sigma_{k+1}+\nu \sum_{k=1}^m
\sigma_k\sigma'_k)$.
The constant $\nu$ is the product of the coupling constant of the Ising model taken here
to be isotropic, with the inverse temperature measured in units that make the Boltzmann
constant unity. We shall refer to $\nu$ as the inverse temperature. 
The critical temperature is defined by $\sinh 2\nu_c=1$.
When the Ising model is on a torus, there is some freedom in the choice of the transfer matrix.
Here this choice is unique as the transfer matrix must include the Boltzmann weights attached to the bonds between the $m$ sites at the boundary and the bonds that tie them to the first inner circle. 

Note that all matrix elements are positive. By Perron-Frobenius theorem, the transfer matrix has real eigenvalues. Moreover its largest eigenvalue is non-degenerate. Let $\omega$ be a non-zero
eigenvector corresponding to this eigenvalue and let $c^s(k_1,k_2,\dots,k_{2n})$
be its components in the state basis. It is known that these components can be chosen such that they are all positive and that their sum is $1$. With this choice, the probability of a given state $\sigma$ 
\begin{equation}\label{eq:proba}
{\text{Pr}}(\sigma=(s;k_1,k_2,\dots,k_{2n}))=c^s(k_1,k_2,\dots,k_{2n})/\kappa
\end{equation}
where 
$$\kappa=\sum_{s\in\{+1,-1\}}\sum_{n=0}^{m/2} \sum_{1\le k_1<k_2< ... <k_{2n}\le m}
c^s(k_1,k_2,\dots,k_{2n}).$$
Note that the $c^s(k_1,k_2,\dots,k_{2n})$'s do not depend on $s$ and this super-index will
be deleted when it is appropriate. 

We use Thompson's notation \cite{thompson} in the
rewriting of $T$ in terms of tensorial blocks. The Pauli matrices 
$\tau^1=\left(\begin{smallmatrix} 0 & 1\\
                              1 & 0 \end{smallmatrix}\right), 
  \tau^2=\left(\begin{smallmatrix}   0 & -i\\
                              i & 0 \end{smallmatrix}\right),
 \tau^3=\left(\begin{smallmatrix}   1 & 0\\
                              0 & -1 \end{smallmatrix}\right),
$
and $\mathbf{1}=\left(\begin{smallmatrix}1 & 0 \\ 0 & 1\end{smallmatrix}\right)$ are used to define 
$2^m\times 2^m$-matrices
$\tau^i_k=\mathbf{1}\otimes \dots \otimes \tau^i
\otimes\dots\otimes\mathbf{1}, i\in\{1,2,3\}$ and $ 1\le k\le m$, where the only non-trivial factor is at position $k$.
The following operators ($\mathbb{C}^{2^m}\rightarrow\mathbb{C}^{2^m}$) will also be used:
$\rho_k  = \tau_1^1 \tau_2^1 \dots \tau_{k-1}^1\tau_{k}^2,$ and $\pi_k  = \tau_1^1 \tau_2^1 \dots \tau_{k-1}^1\tau_{k}^3$ where again $k\in\{1,2,\dots, m\}$. The linear combinations
$a_k = {\textstyle{\frac12}}(\rho_k+i\pi_k)$ and $ a_k^\dagger= {\textstyle{\frac12}} (\rho_k - i\pi_k)$, $k\in\{1,2,\dots, m\}$, satisfy
$a_ka_{k'}^\dagger + a_{k'}^\dagger a_k = \delta_{kk'}\mathbf{1}_{2^m}$ and $ a_k a_{k'} + a_{k'} a_k = \mathbf{0}_{2^m}$. Finally their Fourier coefficients are given by
$$\eta_q=\frac{e^{-i\pi/4}}{\sqrt{m}} \sum_{1\le k \le m} e^{-i q k}a_k\qquad\text{and}\qquad
\eta_q^\dagger = \frac{e^{i\pi/4}}{\sqrt{m}} \sum_{1\le k \le m} e^{i q k}a_k^\dagger$$
for $q\in Q_m$, the set of phases of the $m$-roots of $-1$:
$Q_m=\left\{ \frac{(2j-1)\pi}m, -\frac m2+1\le j\le \frac m2\right\}$.
Using this notation, the transfer matrix is
$$T=(2{\text{\rm sinh}} 2\nu)^{\frac m2} \exp(i\nu \pi_1\rho_m P)
\exp\left(-i\nu\sum_{1\le k \le m-1}\pi_{k+1}\rho_k\right)
\exp\left(i\nu^* \sum_{1\le k\le m} \pi_k\rho_k\right)$$
where $P=\prod_{1\le k\le m}\tau^1_k$, $P^2=\mathbf{1}_{2^m}$, is the operator flipping a
configuration $\sigma=(s;k_i)$ into $(-s; k_i)$ and $\nu^*$ is defined implicitly
by $\sinh 2\nu \sinh 2\nu^*=1$.
Since the Boltzmann weights are invariant 
under the flip of all spins, $P$ commutes with the transfer matrix and $T$ and $P$ can
be diagonalized simultaneously.
The eigenvector $\omega$ belongs to the $+$-eigensubspace of $P$.

The eigenvectors of $T$ are 
constructed with creation and annihilation operators
$\xi_q^\dagger$ and
$\xi_q, q\in Q_m$, obtained from the $\eta_q$ and $\eta_q^\dagger$ by orthogonal
transformations. They are
$\xi_q = \eta_q \cos \phi_q + \eta_{-q}^\dagger \sin\phi_q$ and $\xi_{-q} = \eta_{-q} \cos\phi_q-\eta_{q}^\dagger\sin\phi_q $
with 
$$\tan\phi_q = \frac{(\tanh\, 2\nu+\sinh 2\nu)\sin q}{
1-2\sinh 2\nu\cos q+\sqrt{(\cosh 2\nu-\tanh 2\nu(\cos q+1))(\cosh 2\nu-\tanh 2\nu(\cos q-1))}}.
$$
Though our choice of the transfer matrix, and therefore our expression for $\tan\phi_q$,
are different from those used in
\cite{SML}, the reader will find in that reference the method to obtain this expression.
The eigenspace spanned by $\omega$ is characterized algebraically as the one-dimensional
kernel of the $m$ operators $\xi_q$:
\begin{equation}\label{eq:omega}
\xi_q\omega=0, \qquad q\in Q_m.
\end{equation}
The other eigenstates of $T$ in the maximal subspace $V_+\subset V$ where $P|_{V_+}=1$ are
obtained by acting with an even number of $\xi_q^\dagger$ on the vacuum $\omega$. Because
the $\xi_q$ and $\xi_{q'}^\dagger$ anticommute like the pairs $a_k, a_{k'}^\dagger$'s and $\eta_q,
\eta_{q'}^\dagger$'s, a vector $\xi_{q_1}^\dagger\xi_{q_2}^\dagger\dots\xi_{q_{2i}}^\dagger
\omega$ is a non-zero eigenstate only if the $q_i$'s are distinct. Equation (\ref{eq:omega})
is the one to be solved. 

\subsection{The discrete and continuous cases $n=1$}

Suppose $\sigma$ describes a configuration with a single stretch of $-$-spins and a single
stretch of $+$-spins. Then $\sigma=(s;k_1,k_2)$ and $n=1$. This paragraph is devoted to computing
$\text{Pr}(\sigma=(s;k_1,k_2))$
in the discrete case, cf. equation \eqref{eq:c}, and in the limit, cf. Proposition \ref{prop:conti}.

By translation invariance, it is sufficient to compute $c^+(1,k)$.
The operators $\tau_k^1$ flip the spin at position $k$ and the operators $\rho_k$ and $\pi_{k+1}$
flip all the spins from $1$ to $k$ inclusively. The operators $\xi_q$ are linear combinations of these
and the component along $\sigma_\uparrow$ (the configuration with only $+$-spins) in $\xi_q\omega$
can therefore originate 
only from the action of $\xi_q$ on $\sigma_\uparrow$ or on a $\sigma$ of the
form
$(-;1,k)$. If we denote by 
$(u)_v$ the component along $v$ in the vector $u$ (in the basis given
by the $\sigma$'s), then
\begin{align*}
(\eta_q\omega)_{\sigma_\uparrow} & = \frac{i e^{-i\pi/4}}{2\sqrt m}\left(
     e^{-iq}c_\uparrow - e^{-iqm}c_\downarrow + \sum_{2\le k \le m} e^{-iqk}
c^-(1,k)(1-e^{iq})\right)\\
\intertext{and}
(\eta_{-q}^\dagger\omega)_{\sigma_\uparrow} & = \frac{-i e^{i\pi/4}}{2\sqrt m}\left(
     e^{-iq}c_\uparrow + e^{-iqm}c_\downarrow + \sum_{2\le k \le m} e^{-iqk}
c^-(1,k)(1+e^{iq})\right)
\end{align*}
Using $c_\uparrow=c_\downarrow$ and $e^{iqm}=-1$ for $q\in Q_m$, the vanishing of $(\xi_q\omega)_{
\sigma_\uparrow}$ gives
$i\sum_{2\le k\le m} e^{-iq(k-1)}c^-(1,k)=c_\uparrow \cot (\phi_q+{\textstyle{\frac q2}})$.
Using the discrete inverse Fourier transform and translation symmetry, one gets:
\begin{equation}\label{eq:c}
c^s(k_1,k_2)=-\frac{ic_\uparrow}m\sum_{q\in Q_m}e^{iq(k_2-k_1)}\cot
(\phi_q+{\textstyle{\frac q2}}).
\end{equation}
That $c^s(k_1,k_2)$ is a real number follows from $c^\pm(1,k)=c^\pm(1,m+2-k)$
and the fact that $\phi_q$ is an odd function of $q$. The expression for $\cot (\phi_q+\frac q2)$ is surprisingly similar to that of $\tan \phi_q$:
\begin{align}\label{eq:cot}
d(\nu,q) &=
\cot(\phi_q+{\textstyle{\frac q2}})\notag\\
&=\frac{(1-\tanh 2 \nu)\sin q}{
\sinh 2\nu - \cos q +\sqrt{(\cosh 2\nu-\tanh 2\nu(\cos q+1))(\cosh 2\nu-\tanh 2\nu(\cos q-1))}}.
\end{align} 
We note as a curiosity the following simple relation: $\cot(\phi_q+\frac q2)=\frac{\sqrt2-1}{\sqrt2+1}\tan \phi_q$ at $\nu=\nu_c$. 
We gather in a lemma the elementary properties of $d(\nu,q)$ that are needed for the  limit.\

\begin{Lem}\label{prop:lemma1}
 The function $d:\mathbb{R}^+\times\mathbb{R}\rightarrow\mathbb{R}$ defined by $d(\nu,q)=
\cot (\phi_q+q/2)$ given above has the following properties:
\begin{itemize}
\item[{\em (i)}] for any $\nu\in (0,\infty)$, $d$ has a simple zero at $q=n\pi$, $n$ odd;
\item[{\em (ii)}] around $q=0$ (or any $q=2n\pi, n\in\mathbb{Z}$), the function has the following
behavior
\begin{tabular}{l l}
$\nu>\nu_c$ (subcritical) & $d(\nu,q=0)=0$ and is smooth around this point \\
$\nu=\nu_c$ (critical)    & $d(\nu,q)$ has a jump at $q=0$ with \\
                          & $\sqrt2-1=\lim_{q\rightarrow 0^+}d(\nu_c,q)=
                            -\lim_{q\rightarrow 0^-}d(\nu_c,q)$\\
$\nu<\nu_c$ (supercritical) & $d(\nu,q)$ has a simple pole at $q=0$;
\end{tabular}
\item[{\em (iii)}] outside the behavior along the lines $q=2n\pi$ stated in {\em (ii)}, 
$d(\nu,q)$ is
analytic in $q$ for all inverse temperatures $\nu$.
\end{itemize}
\end{Lem}

\begin{figure}[h!]\label{fig:cotphi}
\begin{center}\leavevmode
\includegraphics[bb = 50 160 670 550,clip,width=0.8\textwidth]{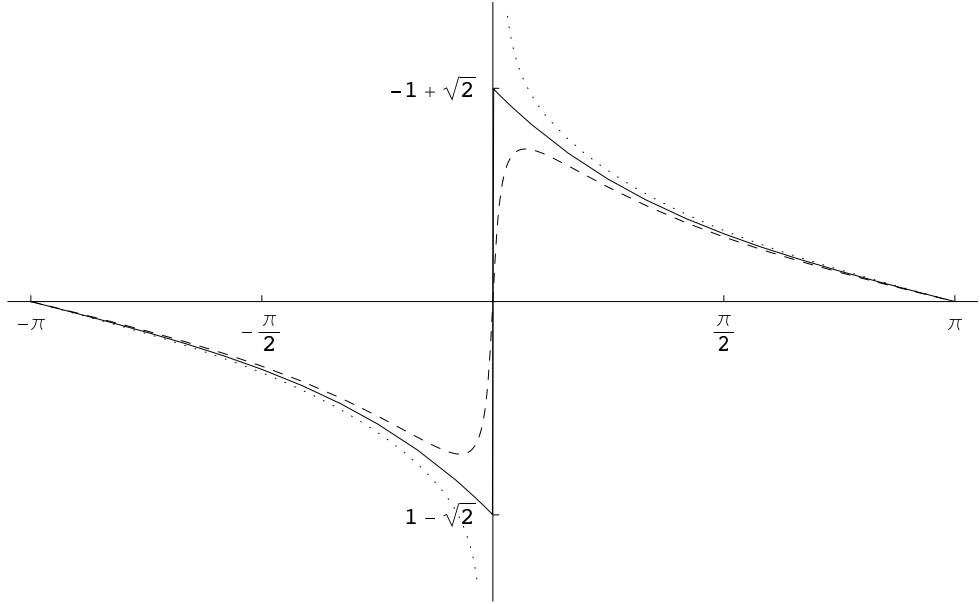}
\end{center}
\caption{The three regimes of the function $\cot(\phi_q+q/2)$: subcritical (dashed curve),
critical (plain curve) and supercritical (dotted curve).}
\end{figure}

Figure \ref{fig:cotphi} draws a graph of these three regimes.

\bigskip
\noindent {\scshape Proof:} 
Property {\em (i)} is obtained by direct evaluation. For {\em (ii)} and {\em (iii)} 
note first that the argument of the square root appearing in the denominator vanishes only when
$\frac{\cosh 2\nu}{\tanh2\nu}=\cos q\pm 1$.
Because the lhs has a single minimum at $\nu=\nu_c=\frac12 
\text{arcsinh}\, 1$ and is equal to $2$,
the argument of the square root is zero for $\nu=\nu_c$ and $q\in2\pi\mathbb{Z}$ only.

For $\nu>\nu_c$, the
square root is positive since $\sinh2\nu>1\ge \cos q$, and the denominator never vanishes. The analyticity of the
subcritical case follows. 
For $\nu\le \nu_c$, the extrema of the argument of the square root can be
found to be at $q\in\pi\mathbb{Z}$ with minima at $q=2n\pi$. The denominator
at $q=n\pi, n$ odd, is equal to 
$1+\sinh2\nu+\sqrt{\cosh2\nu(\cosh2\nu+2\tanh2\nu)}$
 and is therefore
positive. To check whether it vanishes it is sufficient to evaluate it at its minimum $q=0$. At this
point the denominator is $\sinh2\nu - 1 +\sqrt{(1-\sinh2\nu)^2}$. Since $\sinh2\nu_c=1$, we have
that the denominator vanishes only when $q=0$ and $\nu\le \nu_c$. 
For $\nu<\nu_c$, the expansion around $q=0$ of the denominator is 
$\frac{q^2}2(\cosh^22\nu(1-\sinh2\nu))^{-1}+\mathcal{O}(q^4)$.
Because of the $\sin q$ in the numerator, the supercritical behavior described in {\em (ii)} and {\em (iii)}
is established. For $\nu=\nu_c$, the function $d$ takes the simple form
$$d(\nu_c,q)=\frac{(\sqrt2-1)\sin q}{\sqrt2(1-\cos q)+\sqrt{(1-\cos q)(3-\cos q)}}.$$
The leading behavior of its denominator around $q=0$ is
$\sqrt{q^2+\mathcal{O}(q^4)}+\frac{q^2}{\sqrt{2}}+\mathcal{O}(q^4)$
which accounts for the jump. \hfill$\Box$

\medskip

Before turning to the continuum result, we need a technical lemma that rewrites the limit of equation \eqref{eq:c}
when $d$ is smooth.
Equation \eqref{eq:c} shows that $c^s(k_1,k_2)$ depends only on the
ratio $(k_2-k_1)/m$ since $iq(k_2-k_1)=i(2j-1)\pi (k_2-k_1)/m$. 
The meanfingful thing to do is to obtain the limit of that expression
when $\frac{\theta}{2\pi}=\frac{k_2-k_1}{m}\in{\mathbb Q}\cap (0,1)$ is held fixed.
For this fixed ratio, the $m$'s involved in the limit will
have to verify two conditions:
(1) If $\frac{\theta}{2\pi}\in
\mathbb Q$ is written as $\frac pn$ for two relatively prime integers,
then $n$ must divide the $m$'s used in the sequence. 
This ensures that $\frac{\theta}{2\pi}$ is actually of the form $(k_2-k_1)/m$.  
The second condition is more technical and its meaning will be clear in Section \ref{sect:TriAndHexa}. 
In the case of the square lattice, the momenta $q$ can be restricted to the interval $[-\pi,\pi]$. 
For the triangular and hexagonal lattices, they 
lie in $[-\frac{\pi}2,\frac{\pi}2]$.
We introduce $\gamma\in\mathbb N$ such that the momentum range is $[-\frac{2\pi}\gamma,\frac{2\pi}\gamma
]$. The second condition on $m$'s is: (2) $r:=\frac m\gamma \mod n$ is constant.

\begin{Lem} \label{prop:lemma2} (i)
Let $z:[0,\frac{2\pi}\gamma]\rightarrow\mathbb R$, $\gamma\in\mathbb N$
be such that $z,z'$ and $z''$ exist and are continuous on $[0,\frac{2\pi}\gamma]$.
Let $\frac{\theta}{2\pi}=\frac pn$ with $\gcd(p,n)=1$. Then
\begin{align}
\lim_{m\rightarrow\infty}
     2\sum_{j=1}^{m/\gamma} \sin\Big((2j-1)\frac{\theta}2\Big)
     z\left((2j-1)\frac{\pi}m\right)& = \frac1{\sin (\theta/2)}\left(
     z(0)-z(\textstyle{\frac{2\pi}\gamma})\cos (r\theta)\right)\\
\lim_{m\rightarrow\infty}
     2\sum_{j=1}^{m/\gamma} \cos\Big((2j-1)\frac{\theta}2\Big)
     z\left((2j-1)\frac{\pi}m\right)& = \frac{\sin(r\theta)}{\sin(\theta/2)}
     z(\textstyle{\frac{2\pi}\gamma})
\end{align}
where the limits are taken over integer $m$'s such that $n$ and $\gamma$ divide 
$m$ and such that $r:=m/\gamma \mod n$ is constant.

\noindent (ii) If $z(\frac{2\pi}\gamma)=0$, the above results hold without the
requirement that $r$ be held constant.
\end{Lem}

\noindent{\scshape Proof:} {\em (i)} We divide the range of $j\in \{ 1, 2, \dots,
\frac m\gamma\}$ into subsets of $n$ consecutive elements: $\{1,2,\dots, n\},
\allowbreak \{n+1, n+2, \dots, 2n\}$ and so on. 
There will be $r$ elements left.
Omitting these $r$ terms for the present, we consider the sum 
\begin{align}
S_m&:= 
 \sum_{l=0}^{[\frac m{n\gamma}]-1} \sum_{j=0}^{n-1}
      e^{i(2(j+ln)+1)\theta/2} z((2(j+ln)+1)\pi/m)
   &= \sum_{l=0}^{[\frac m{n\gamma}]-1} \sum_{j=0}^{n-1}
      e^{i(2j+1)\theta/2} z(q_l+2j\pi/m)
\label{eq:lp1}
\end{align}
where $q_l=(2ln+1)\pi/m$ and $[x]$ stands for the integer part of $x$. 
By Taylor expansion $z(q_l+2j\pi/m)=z(q_l)+\frac{2j\pi}m z'(q_l)+\frac{4j^2\pi^2}{2m^2} \epsilon_{l,j}$
with $|\epsilon_{l,j}|\leq s:=\sup_{q\in [0,2\pi/\gamma]}
|z''(q)|
<\infty$.
The two sums
$\sum_{j=0}^{n-1} e^{i(2j+1)\theta/2}$ and $
\sum_{j=0}^{n-1} je^{i(2j+1)\theta/2}$
are easily calculated. At $\frac{\theta}{2\pi}=\frac pn$, the first vanishes
and the second is $\frac n{2i\sin \theta/2}$.
The sum (\ref{eq:lp1}) is then 
\begin{align*}
\sum_{l=0}^{[\frac m{n\gamma}]-1} \sum_{j=0}^{n-1} e^{i(2j+1)\theta/2} &
  \left(z(q_l)+\frac {2j\pi}m(z'(q_l)+\epsilon_{l,j})\right) \\
  &= \frac{n\pi}{im}\frac1{\sin \theta/2} \sum_{l=0}^{[\frac m{n\gamma}]-1} z'(q_l)
    + \frac{2\pi^2}{m^2} \sum_{l=0}^{[\frac m{n\gamma}]-1} \sum_{j=0}^{n-1} 
    j^2 e^{i(2j+1)\theta/2} \epsilon_{l,j}.
\end{align*}
The second sum goes to zero since its absolute value is smaller than $2\pi^2n^2s/\gamma m.$
The first sum is a Riemann sum whose limit on $m$ is
\begin{equation}
\lim_{m\rightarrow \infty} S_m=\frac1{2i\sin\theta/2}\int_0^{2\pi/\gamma}
z'(q)dq = \frac i{2\sin\theta/2}(z(0)-z(\textstyle{\frac{2\pi}\gamma})).
\label{eq:lp2}
\end{equation}

We now turn to the $r$ residual terms:
$R_m:=\sum_{j=m/\gamma-r+1}^{m/\gamma}e^{i(2j-1)\theta/2}z((2j-1)\pi/m)$.
We expand $z$ around $\frac {2\pi}\gamma$:
$z((2j-1)\pi/m)=z(2\pi/\gamma)+z'(x_j)
\left(\frac{(2j-1)\pi}m-\frac{2\pi}\gamma\right)$
for some                                     
$x_j\in \left[\frac{(2j-1)\pi}m,\frac{2\pi}\gamma\right].$
The remainder $R_m$ is
\begin{equation}
R_m=z(\textstyle{\frac{2\pi}\gamma})\sum_{j=m/\gamma-r+1}^{m/\gamma}e^{i(2j-1)\theta/2}
+\sum_{j=m/\gamma-r+1}^{m/\gamma}e^{i(2j-1)\theta/2}\left( \frac{2\pi}\gamma
-\frac{(2j-1)\pi}m\right)z'(x_j).\label{eq:lp4}
\end{equation}
The second sum vanishes in the limit because
its absolute value is smaller than $\frac{(2r-1)\pi}m \sup |z'(q)|$.
The first sum is easily calculated and
\begin{equation}
\lim_{m\rightarrow\infty} R_m=z({\textstyle{\frac{2\pi}\gamma}})e^{ir\theta/2}
\frac{\sin(r\theta/2)}{\sin(\theta/2)}.\label{eq:lp3}
\end{equation}
The result follows by taking the real and imaginary parts of $\lim (S_m+R_m)$
as given by (\ref{eq:lp2}) and (\ref{eq:lp3}).

\noindent {\em (ii)} The result follows from (\ref{eq:lp4}) where, again, the 
second sum goes to zero and the factor $z(\frac{2\pi}\gamma)=0$ removes the 
$r$-dependent sum. \hfill$\Box$

\begin{Prop}[Continuous case]\label{prop:conti}
Set $\theta=\frac{2\pi k}m$ and denote by $\lim$ the process of
taking the limit $k,m\rightarrow \infty$ while keeping $\theta$ fixed. The
thermodynamical limits of
$c^s(k_1,k_2)/c_\uparrow$ (eq.{} (\ref{eq:c})) 
with $k=k_2-k_1>0$ are
\begin{itemize}
\item[{\em (i)}] {\em (supercritical, $\nu<\nu_c$)}
$$\lim -\frac im\sum_{q\in Q_m} e^{iqk} d(\nu,q)=(1-\text{\rm tanh\ }2\nu)(1-\text{\rm sinh\ }2\nu)
\text{\rm cosh}^2\ 2\nu,$$
independent of $\theta$;
\item[{\em (ii)}] {\em (critical, $\nu=\nu_c$)}
$$\lim -i\sum_{q\in Q_m} e^{iqk} d(\nu_c,q)=\frac{\sqrt2-1}{\sin\theta/2};$$
\item[{\em (iii)}] {\em (subcritical, $\nu>\nu_c$)} 
$$\lim -i\sum_{q\in Q_m}e^{iqk} d(\nu,q)$$
goes to a Dirac distribution in the following sense: if $f:{\mathbb T}^1\rightarrow {\mathbb R}$
is a continuous function on the circle, then
$$\lim_{m\rightarrow \infty} -i\frac{2\pi}m \sum_{k=1}^{m-1}f\left(\frac{2\pi k}m\right)
\sum_{q\in Q_m} e^{iqk} d(\nu,q)=\gamma f(0)$$
with
$$\gamma=2\int_0^\pi d(\nu,q)\cot \frac q2\, dq.$$
\end{itemize}
\end{Prop} 
We stress that only the properties stated in Lemma \ref{prop:lemma1}
are used in the proof and that it is the behavior {\em (ii)} at $q=0$ that
decides between the three regimes.
\medskip

\noindent{\scshape Proof:} {\em (critical)} Note that 
$\lim -i\sum_{q\in Q_m}e^{iqk}d(\nu_c,q) = \lim 2\sum_{q\in Q_m^+}d(\nu_c,q)\sin kq$. 
By Lemma \ref{prop:lemma1}, $d(\nu_c,0^+)=\sqrt2-1$ and $d(\nu_c,\pi)=0$ and
$d(\nu_c,q)$ is analytic in $q$ for $q\in[0,\pi]$. Therefore the limit follows
from Lemma \ref{prop:lemma2} {\em (ii)}:
$\lim -i\sum_{q\in Q_m} e^{iqk}d(\nu_c,q)=\frac{\sqrt2-1}{\sin\theta/2}$.

\medskip

\noindent {\em (supercritical)\/} To prove the supercritical case, we first replace the function
$d(\nu,q)$ in the limit by $\frac 1q$. Then
$-\frac im\sum_{q\in Q_m} e^{iqk}\frac1q=\frac1\pi\sum_{j=1}^{m/2}\frac{\sin\theta(j-\frac12)}{
j-\frac12}$.
This sum goes to $\frac{\pi}2$ when $m\rightarrow\infty$ for any value of $\theta\in(0,2\pi)$.
(See \cite{Encyclo}, App.{} A, Table II.) Therefore
$\lim -\frac im\sum_{q\in Q_m} e^{iqk}\frac1q=\frac12$.
Let us write $d(\nu,q)$ as
$d(\nu,q)=\frac{\alpha}q-\frac{\alpha}{\pi^2}q+g(q)$
where $\alpha$ is the residue of $d(\nu,q)$ at $q=0$:
$\alpha=2 (1-\text{\rm tanh\ }2\nu)(1-\text{sinh\ }2\nu)\,
\text{cosh}^2\ 2\nu$.
Then the new function $g$ is analytic on $[0,\pi]$ with $g(0)=g(\pi)=0$. 
This function will therefore not contribute to the limit by Lemma \ref{prop:lemma2}. 
(Note that, in the present case, there is even a supplementary factor
$\frac1m$ in the limit.) We thus obtain
$$\lim_{m\rightarrow \infty} -\frac im\sum_{q\in Q_m} e^{iqk}d(\nu,q)=\frac{\alpha}2-\frac{\alpha}{
\pi^2}\lim_{m\rightarrow\infty} -\frac im\sum_{q\in Q_m}q e^{iqk}.$$
Because $|\sum_{j=1}^{m/2}\sin \theta(j-\frac12)|\le 1/|\sin (\theta/2)|$ and
$\sum_{j=1}^{m/2} j\sin\theta(j-\frac12)$ grows at most as $m$ times a constant 
(depending on $\theta$) as $m\rightarrow \infty$, the remaining limit on the right-hand side
vanishes. The supercritical case follows.

\medskip

\noindent {\em (subcritical)\/}
The limit under consideration now is
$\lim_{m\rightarrow\infty} -\frac{2\pi i}m\sum_{k=1}^{m-1}
f\left(\frac{2\pi k}m\right) \sum_{q\in Q_m} e^{iqk}d(\nu,q)$
for a continuous function $f$ on the circle $\mathbb T^1$. The case of a constant function $f$
is simple:
\begin{align*}
\lim_{m\rightarrow\infty} -\frac{2\pi i}m\sum_{k=1}^{m-1}f(0) \sum_{q\in Q_m} e^{iqk}d(\nu,q)
&=f(0) \lim_{m\rightarrow \infty} -\frac{2\pi i}m \sum_{q\in Q_m} d(\nu,q)\sum_{k=1}^{m-1} e^{iqk}\\
&= f(0) \lim_{m\rightarrow\infty} \frac{2\pi}m\sum_{q\in Q_m}d(\nu,q)\cot {\textstyle{\frac q2}}
\end{align*}
and, because $d(\nu,q)$ has a simple zero at $q=0$, the summand is bounded and the limit
is $2 f(0) \int_0^\pi d(\nu,q)\cot {\textstyle{\frac q2}} dq$.

We shall now concentrate on the continuous function $g:{\mathbb T}^1\rightarrow\mathbb R$
given by $g(\theta)=f(\theta)-f(0)$ that vanishes at $\theta=0$:
\begin{equation}\label{eq:sub}
\lim_{m\rightarrow\infty}-\frac{2\pi i}m\sum_{k=1}^{m-1}g\left(\frac{2\pi k}m\right)
\sum_{q\in Q_m}e^{iqk}d(\nu,q).
\end{equation}
First note that the inner sum is an approximation of the Fourier coefficients of
$d(\nu,q)$
$$\frac1m\sum_{q\in Q_m} e^{iqk}d(\nu,q)=\frac1{2\pi}\int_{-\pi}^\pi e^{iqk}d(\nu,q)dq+
\frac{\text{constant}}{m^2}\frac{\partial^2d}{\partial q^2}(\nu,q')$$
for some $q'\in[-\pi,\pi]$. With the factor $\frac 1{m^2}$, the correction terms disappear upon
taking the limit. Therefore the limit (\ref{eq:sub}) is
$\lim_{m\rightarrow\infty} -2\pi i\sum_{k=0}^{m-1} g\left(\frac{2\pi k}m\right) d_{-k}$
with
$d_k=\frac1{2\pi}\int_{-\pi}^\pi e^{-iqk} d(\nu,q)dq$.
Because $d(\nu,q)$ is real-analytic, its Fourier coefficients decrease exponentially, i.e.{} there
exist $c_1$ and $c_2>0$ such that 
$|d_k|<c_1 e^{-c_2|k|}$, for $k\in\mathbb Z$.
(See, for example, V.16 of \cite{Godement}.)
Let $\epsilon>0$ and 
$M= \max_{\theta\in\mathbb T^1} \left|g(\theta)\right|$. Then there exists
$K\in\mathbb N$ such that
$\left|\sum_{k=K}^{m-1}g\left(\frac{2\pi k}m\right)d_{-k}\right|\le c_1 M\frac{e^{-c_2
K}}{1-e^{-c_2}} <\frac{\epsilon}2$  for $m>K$.
Let $D=\max\{|d_{-1}|,|d_{-2}|,\dots,|d_{-(K-1)}|\}$. The remaining terms are bounded by
$\left| \sum_{k=1}^{K-1} g\left(\frac{2\pi k}m\right) d_{-k}\right|< DK\max_{\theta\in[0,
\frac{2\pi K}m]}|g(\theta)|$.
Because $g$ is continuous and $g(0)=0$, there must be a $N>K$ such that, if $m>N$, then 
$DK\sup_{\theta\in[0,
\frac{2\pi K}m]}|g(\theta)|<\frac{\epsilon}2$.
Therefore the limit (\ref{eq:sub}) vanishes and 
$$\lim_{m\rightarrow\infty} -\frac{2\pi i}m\sum_{k=1}^{m-1} f\left(\frac{2\pi k}m\right)
\sum_{q\in Q_m} e^{iqk}d(\nu,q)=\gamma f(0)\qquad{\text{\rm with}}\qquad
\gamma=2\int_0^\pi d(\nu,q)\cot {\textstyle{\frac q2}}dq.$$
\hfill $\Box$

\subsection{The discrete and continuous cases for $n<\frac m2$}\label{sec:ngeneral}

We now turn to the general case when the number of spinflips is any
number between $0$ and $m$. The calculation for $n=2$ generalizes trivially
to the general case and we give more details for this case. We drop the
superindex ``$s$'' on $c^s(k_1, k_2, \dots, k_{2n})$ as it does not play
any role here.

To mimick the argument of the previous section, we shall write down the
component, in $\xi_q\omega$ (see equation \eqref{eq:omega}), of the vector $v$ 
\begin{center}
\begin{tabular}{ccccccccccccc}
$\qquad 1$ & 2 & 3 &$\dots$ &$ k_1-1$ & $k_1$ & $k_1+1$ & $\dots$ & $k_2-1$ &
$k_2$ &
$k_2+1$ & $\dots$ & $m\ \ $ \\
$v=(\ -$ & $-$ & $-$ & $\dots$ & $-$     & $+$   &  $+$    & $\dots$ & $+$ &
$-$   & $-$     & $\dots$ & $-\ )$.
\end{tabular}
\end{center}
Because the operators $\xi_q$'s (and $\eta_q$'s) are made of the operators
$\rho_k$ and $\pi_k$ that flip all spins to the left of the site $k$ and
may or may not change the spin $k$, the only vectors that may contribute to
$(\xi_q\omega)_v$ are the seven following (families of) configurations:
\begin{center}
\begin{tabular}{l|lcc|lcc|lcr|}
       & $1$ \hfill  &         &       & $k_1$\hfill &         &       &
$k_2$\hfill &         &   $m$     \\ 
I\qquad\ & $- - - - -$ & $\dots$ & $- -$ & $+ + + + +$ & $\dots$ & $+ +$ & 
$- - - - -$ & $\dots$ & $- -$  \\
II\qquad\ & $+ + + - -$ & $\dots$ & $- -$ & $+ + + + +$ & $\dots$ & $+ +$ & 
$- - - - -$ & $\dots$ & $- -$  \\
III\qquad\ & $+ + + + +$ & $\dots$ & $+ +$ & $+ + + + +$ & $\dots$ & $+ +$ & 
$- - - - -$ & $\dots$ & $- -$  \\
IV\qquad\ & $+ + + + +$ & $\dots$ & $+ +$ & $- - - + +$ & $\dots$ & $+ +$ & 
$- - - - -$ & $\dots$ & $- -$  \\
V\qquad\ & $+ + + + +$ & $\dots$ & $+ +$ & $- - - - -$ & $\dots$ & $- -$ & 
$- - - - -$ & $\dots$ & $- -$  \\
VI\qquad\ & $+ + + + +$ & $\dots$ & $+ +$ & $- - - - -$ & $\dots$ & $- -$ & 
$+ + + - -$ & $\dots$ & $- -$  \\
VII\qquad\ & $+ + + + +$ & $\dots$ & $+ +$ & $- - - - -$ & $\dots$ & $- -$ & 
$+ + + + +$ & $\dots$ & $+ +$  \\
\end{tabular}
\end{center}
The configurations I, III, V and VII have 2 spinflips, the others have 4. The
action of $\eta_q$ on these terms in $\omega$ is given by $\frac12ie^{-i\pi/4}
m^{-1/2}$ times
\begin{center}
\begin{tabular}{ll}
I \qquad\   & ${\displaystyle{-e^{-iq}c(k_1,k_2)}}$ \\
II\qquad\  & ${\displaystyle{\sum_{l=2}^{k_1-1}(e^{-iq(l-1)} - e^{-iql}) c(l,
k_1, k_2, m+1)}}$ \\
III\qquad\ & ${\displaystyle{(e^{-iq(k_1-1)}+e^{-iqk_1})c(k_2,m+1) }}$ \\
IV\qquad\ & ${\displaystyle{\sum_{l=k_1+1}^{k_2-1}(-e^{-iq(l-1)}+e^{-iql})
c(k_1,l,k_2,m+1) }}$\\
V\qquad\ & ${\displaystyle{ (-e^{-iq(k_2-1)} - e^{-i q k_2}) c(k_1,m+1)}}$ \\
VI\qquad\ & ${\displaystyle{\sum_{l=k_2+1}^m(e^{-iq(l-1)}-e^{-iql})
c(k_1,k_2,l,m+1) }}$\\
VII\qquad\ & ${\displaystyle{e^{-iqm}c(k_1,k_2) }}$\\
\end{tabular}
\end{center}
Similar expressions can be obtained for the action of $\eta_{-q}^\dagger$.
We shall write the component of $v$ in the equation $\xi_q\omega=
(\eta_q\cos \phi_q+\eta_{-q}^\dagger\sin\phi_q)\omega=0$ in the form
$$\underset{\text{4 spinflips}}{\underbrace{\text{II $+$ IV $+$ VI}}} = - \big(\ 
\underset{\text{2 spinflips}}{\underbrace{\text{$($I $+$ VII $)$ $+$ III $+$ V}}}\ \big) .$$
The left-hand side is
$$
ie^{i\pi/4}e^{iq/2}  \Big( \underset{\text{II}}{\underbrace{+\sum_{l=2}^{k_1-1}}}\ \ 
  \underset{\text{IV}}{\underbrace{-\sum_{l=k_1+1}^{k_2-1}}}\ \ 
  \underset{\text{VI}}{\underbrace{+\sum_{l=k_2+1}^m}}
\Big) 
e^{-iql}\sin(\phi_q+q/2)c(\pi(l,k_1,k_2,m+1))$$
where the symbol $\pi$ appearing in $c(\pi(l,k_1,k_2,m+1))$ is the permutation that orders
the integers $l, k_1, k_2, m+1$. One can rewrite compactly this expression by setting
$c(k_1,k_2,k_3,k_4)=0$ whenever two of the arguments coincide and by denoting the parity of the
permutation $\pi$ by $(-1)^{l(\pi)}$:
\begin{equation}\label{eq:lhs}
ie^{i\pi/4}e^{iq/2} \sum_{l=2}^m (-1)^{l(\pi)} e^{-iql}\sin (\phi_q+q/2) c(\pi(l,k_1,k_2,m+1)).
\end{equation}
The right-hand side is 
\begin{equation}\label{eq:rhs}
e^{i\pi/4}e^{iq/2}\cos(\phi_q+q/2) \big(  \underset{\text{$($ I $+$ VII $)$}}{\underbrace{e^{-iq}c(k_1,k_2)}}
\ \    \underset{\text{III}}{\underbrace{- e^{-iqk_1}c(k_2,m+1)}}
\ \    \underset{\text{V}}{\underbrace{+ e^{-iqk_2}c(k_1,m+1)}}
\big)
\end{equation}
The discrete inverse Fourier transform gives the desired expression
\begin{align*}
(-1)^{l(\pi)} & c(\pi(k,k_1,k_2,m+1))\\
&= \frac im \sum_{q\in Q_m} e^{iqk} d(\nu,q) \left( e^{-iqk_1}c(k_2,m+1)
- e^{-i q k_2} c(k_1,m+1)
+ e^{-iq(m+1)} c(k_1,k_2) \right).
\end{align*}

The generalization to $2n$ spinflips is immediate. It will be obtained by examining
the component of a $(2n-2)$-spinflip configuration in $(\xi_q\omega)$. The contribution of
$(2n)$-spinflip configurations will be of the same form as the left-hand side in
(\ref{eq:lhs}). (Note that the above calculation depends only on the positions $k_1$
and $k_2$ and on the number of stretches of constant signs that are flipped by
the operators $\rho_k$ and $\pi_k$.) Similarly the $(2n-2)$-spinflip configurations
will lead to $(2n-1)$ terms with alternating sign as in (\ref{eq:rhs}). We therefore
get the following recursive formula.

\begin{Prop}[Recursive form] Let the number $2n$ of spinflips be such
that $1\le n\le \frac m2$ and let $1\le k_1 < k_2 < \dots < k_{2n} \le m$ be their
positions. Then
\begin{equation}\label{eq:rec}
c(k_1, k_2, \dots, k_{2n})=\frac im \sum_{q\in Q_m} \sum_{j=2}^{2n}
d(\nu,q) e^{iq(k_1-k_j)}(-1)^j c(\widehat{k_1},k_2, \dots, k_{j-1},\widehat{k_j},k_{j+1},\dots,
k_{2n})
\end{equation}
where $c(\widehat{k_1},k_2,\dots,k_{j-1},\widehat{k_j},
	k_{j+1},\dots,k_{2n})=c(k_2,k_3,\dots,k_{j-1},k_{j+1},\dots,k_{2n})$, if  $n\ge 2$ and $c_\uparrow$ when $n=1$.
\label{prop:rec}
\end{Prop}

We restrict the discussion of the thermodynamical limit to the critical regime.

\begin{Prop}[Continuous case at criticality] Set $\theta_i=2\pi k_i/m, 1\le i\le 2n$ with $n\ge 2$ and denote
by $\lim$ the process of taking the limit $k_1, k_2, \dots, k_{2n}, m\rightarrow \infty$ while
keeping the $\theta_i$'s fixed. The thermodynamical limit of $c(k_1,k_2, \dots, k_{2n})$ is
$$
p(\theta_1,\theta_2,\dots,\theta_{2n})  = \lim \frac{m^n}{c_\uparrow} c(k_1, k_2, \dots, k_{2n})  = \sum_{j=2}^{2n} \frac{(-1)^j(\sqrt2-1)}{\sin(\theta_{j1}/2)} p(\widehat{\theta_1},\theta_2,
\dots, \widehat{\theta_j},\theta_{j+1},\dots, \theta_{2n})$$
where $\theta_{ij}=\theta_i-\theta_j$ and
$p(\widehat{\theta_1},\theta_2,
\dots, \theta_{j-1},\widehat{\theta_j},\theta_{j+1},\dots, \theta_{2n}) =
p(\theta_2,\theta_3, \dots, \theta_{j-1},\theta_{j+1},\dots, \theta_{2n})$ if $n\ge 2$ and $=1$ when  $n=1$.
\label{prop:conti 2n}
\end{Prop}

\noindent{\scshape Proof:} 
This follows from Proposition \ref{prop:conti} for $
\nu=\nu_c$ and the elementary fact that the limit of a product is the product
of the limits as long as these limits exist.\hfill $\Box$

As an example we give the continuous case for $n=2$:
\begin{equation}
p(\theta_1,\theta_2,\theta_3,\theta_4)=(\sqrt2-1)^2\left(
\frac1{\sin\frac12\theta_{12} \sin\frac12\theta_{34} } -
\frac1{\sin\frac12\theta_{13} \sin\frac12\theta_{24} } +
\frac1{\sin\frac12\theta_{14} \sin\frac12\theta_{23} }\right).
\label{eq:prn2}
\end{equation}
This particular case was used in \cite{ArguinYSA} to obtain the probability that the interface, separating the constant-spin clusters meeting at $\theta_1$, ends at $\theta_2$.

%
%

\section{The distribution of the number of
spinflips}\label{sect:flips}
Is there a ``typical'' number of spinflips at the boundary of a long cylinder?
Or more precisely, what is the probability distribution of the random variable 
$$Y_m=\frac nm=\frac{\text{$\#$ spinflips}/2}{\text{$\#$ sites at the boundary}}\in[0,\frac12]$$
as $m\rightarrow \infty$? 
De Coninck \cite{deConinck} proved that the rescaled 
variable $Y_m$ is Gaussian in the limit.
We use the results of the previous section to recover this 
and to determine explicitly the mean and variance.

\begin{Prop} \label{prop limit spinflips}
The random variable $Y_m=n/m$ on the set of configurations at the
extremity of a half-infinite cylinder behaves at criticality as
\begin{equation*}
\lim_{m\rightarrow\infty} \text{\rm Pr}_m\left(\frac{Y_m-\mu}{\sigma/\sqrt{m}}<x\right) = 
\frac1{\sqrt{2\pi}}\int_{-\infty}^x e^{-x^2/2}dx
\end{equation*}
with
$\mu={{\frac12}}-{{\frac1{2\pi}}}(\sqrt{2}+1)$ and $
\sigma^2= {{\frac1{2\pi}}}(7+5\sqrt{2})-{{\frac38}}(3+2\sqrt{2})$.
\end{Prop}

In particular the above proposition asserts that, at the boundary of a long cylinder covered by 
a square lattice, there are on average $2m \ \mu\approx 2m\times 0.115766\dots$ spinflips, that is there is
one at almost every 4 sites! 

Proposition \ref{prop:conti 2n} gives the limit of the discrete probability
${\text Pr}_m(k_1<k_2<\dots<k_{2n} | n)$ for the square lattice. 
Unfortunately the weight included in the limit process (proportional to $m^n$) rules out 
comparing relative probabilities of all configurations with
a given number of spinflips. 
One has therefore to go back to the recursive form of Proposition \ref{prop:rec} for
the probability $c^s(k_1,k_2,\dots,k_{2n})$ for finite $n$ and $m$. 
One may restrict the comparison
to configurations whose spin at position 1 is $+$ and again
drop the superindex $s$. We shall use the notation $d(q)$ for $d(\nu_c,q)$.
We are interested in computing
\begin{equation}\lim_{\begin{subarray}{c}n,m\rightarrow\infty\\ n/m=\kappa {\text{\rm \, fixed}}
\end{subarray}} {\text{\rm Pr}}_m(Y_m=n/m)
\end{equation}
where
${\text{\rm Pr}}_m(Y_m=n/m)=\sum_{1\le k_1<k_2<\dots<k_{2n}\le m} c(k_1,k_2,\dots,k_{2n})$.
The recursive formula (\ref{eq:rec}) can be used to give an explicit expression 
of $c(k_1,k_2,\dots,k_{2n})$ as a sum over certain permutations:
\begin{equation}
\text{Pr}_m(Y_m=n/m) =
 \frac{c_\uparrow}{n!} \sum_{\begin{subarray}{c}q_1, q_2, \dots, q_n\in Q_m^+\\ \text{distinct
$q$'s}\end{subarray}}\ \ 
   \prod_{1\le j \le n} d(q_j) \cot {\textstyle{\frac12}} q_j \label{eq:qm}
\end{equation}
We first prove Proposition \ref{prop limit spinflips} using \eqref{eq:qm}.

\medskip

\noindent{\scshape Proof: }  
We define $f(q):=d(q)\cot(q)$ for short.
It is easily checked that the above expression can be rewritten in the form of the probability of $n$ successes for $m/2$ independent Bernoulli trials, each with probability $p_{m,i}:=f(q_i)/(1+f(q_i))$ so $1-p_{m,i}=1/(1+f(q_i))$. As a result the normalization is $\frac{n!}{c_{\uparrow}}=\prod_{i=1}^{m/2}(1+f(q_i))$. Let us write $\epsilon_{m,i}$ for a Bernouilli variable with probability $p_{m,i}$. 
$Y_m$ is then simply $\frac{1}{m}\sum_i^{m/2}\epsilon_{m,i}$. 
Note that the variables $\epsilon_{m,i}$ form a triangular array. 
(See, for example, \cite{Bi}.)
Moreover they are independent. 
Therefore the central limit theorem can be applied if the Lindeberg condition is verified.
Since the variables $\epsilon_{m,i}$ are uniformly bounded, the condition reduces to verify that the limit of the variance of $\sqrt{m}Y_m$ exists. But we have
$$\sigma^2:=\lim_{m\to\infty}\frac1m\sum_{i=1}^{m/2}p_{m,i}(1-p_{m,i})=\frac1{2\pi}\int_0^\pi\frac{f(q)}{(1+f(q))^2}dq.$$
In addition, the mean of $Y_m$ converges to
$$ \mu:=\lim_{m\to\infty}\frac{1}{m}\sum_{i=1}^{m/2}\epsilon_{m,i}=\frac{1}{2\pi}\int_0^\pi\frac{f(q)}{1+f(q)}dq, $$
The integrands in the expression of $\mu$ and $\sigma$ can be put into
simple forms:
\begin{align*}
\frac{f(q)}{1+f(q)}&=-\frac12(1+\sqrt{2})\left(1-2\sqrt2+\cos q+\sqrt{(1-\cos q)(3-\cos q)}\right)\\
\frac{f(q)}{(1+f(q))^2}&=\frac1{8\pi(3-2\sqrt2)}\left((-2-2\sqrt2)+(4+2\sqrt2)\cos q-2\cos^2 q\right.  \\ 
                       & \qquad \qquad + \left. (2\sqrt2-2\cos q)\sqrt{(1-\cos q)(3-\cos q)}\right).
\end{align*}
Integration can then be done and give
$
\mu={{\frac12}}-{{\frac1{2\pi}}}(\sqrt{2}+1)\approx{}\, 0.115766\dots$  and $
\sigma^2=\left( {{\frac1{2\pi}}}(7+5\sqrt{2})-{{\frac38}}(3+2\sqrt{2})\right)\approx{}\, 0.0538198\dots$\ .\hfill$\Box$

\medskip

We turn now to the proof of \eqref{eq:qm}.
Let $S_{2n}$ denote the permutation groups of $2n$ elements. We shall call {\em pfaffian}
a permutation  $\pi\in S_{2n}$ of the integers $\{1,2,\dots, 2n\}$ that satisfies
\begin{alignat}{14}\label{eq:pfaffian}
\pi(1) & < \pi(2), &\quad &  & \quad  \pi(3) & < \pi(4), & \quad &   &\quad & \dots, & \quad &   &\quad & \pi(2n-1) & < \pi(2n) \notag \\
       & \pi(1)    &      &< &               & \pi(3)    &       & < &      & \dots &       & < &      &           & \pi(2n-1).
\end{alignat}
(These are the conditions on the indices of terms appearing in
the pfaffian of an anti-symmetric $2n\times2n$ matrix. See,
for example, \cite{McCoy,dFMS} or \cite{Feldman}.) The set of
all these permutations $\subset S_{2n}$ will be denoted $\text{Pf}_{2n}$. 
Two well-known facts are useful here.
First suppose that $\pi(1)\neq 1$. Then there exists $i>1$ such that $\pi(i)=1$. If $i$
is even, then $\pi(i-1)<\pi(i)$ and
one of the pfaffian inequalities is surely false.
If $i$ is odd, then it is $\pi(i-2)<\pi(i)$ that is false. Therefore $\pi(1)=1$ for
$\pi\in\text{Pf}_{2n}$. Second we obtain the cardinality of $\text{Pf}_{2n}$. We
have just seen that $\pi(1)=1$. All values $2\le i\le 2n$ are possible choices
for $\pi(2)$. There are $(2n-1)$ of them. By an argument similar to the one
leading to $\pi(1)=1$, one shows that $\pi(3)$ must be the smallest integer
left in $\{1,2,\dots,2n\}$ after deletion of $\pi(1)$ and $\pi(2)$. There are
then $(2n-3)$ choices for $\pi(4)$. Repeating the argument one gets $|\text{Pf}_{2n}|
=(2n-1)!!\ (< |S_{2n}|=(2n)!)$.

Note that the previous argument for the cardinality of $\text{Pf}_{2n}$ follows the
process by which the recursive expression (\ref{eq:rec}) constructs the general term
of $c(k_1, k_2, \dots, k_{2n})$. For example, $c(k_1, k_2,k_3, k_4)$ is given by\begin{equation*}
c_\uparrow\left(\frac im\right)^2\sum_{q_1,q_2\in Q_m} d(q_1) d(q_2)
    \big(    e^{iq_1(k_1-k_2)}e^{iq_2(k_3-k_4)}
                  - e^{iq_1(k_1-k_3)}e^{iq_2(k_2-k_4)}
	              + e^{iq_1(k_1-k_4)}e^{iq_2(k_2-k_3)}\big).
\end{equation*}
The permutations $(\pi(1),\pi(2),\pi(3),\pi(4))$ appearing here are
precisely the three pfaffian permutations of $\text{Pf}_4$. To obtain a similar form for a general
$n$, let us denote the momentum introduced at the $j$-th use of the recursion formula by $q_j$ and 
the two indices on the positions $k$'s by $\pi(2j-1)$ and $\pi(2j)$. At the first use the
formula (\ref{eq:rec}) forces $\pi(1)$ to be $1$ and $\pi(2)$ to be any number between $2$ and $2n$.
This first step gives rise to the $(2n-1)$ terms $(-1)^i e^{iq_1(k_1-k_i)}$. At the second use,
$\pi(3)$ will be the smallest of the remaining integers in $\{1,2,\dots, 2n\}$ and $\pi(4)$ will take
any of the $(2n-3)$ remaining values. It is clear then that the indices $\pi(2j-1)$ and $\pi(2j)$
appearing in $\prod_{1\le j\le n}e^{iq_j(k_{\pi(2j-1)}-k_{\pi(2j)})}$ are those obtained by a pfaffian
permutation of $\{1,2,\dots,2n\}$ and that all such permutations occur precisely once. Because $(i-2)$
is the number of neighbor transpositions necessary to go from $(1,2,\dots, 2n)$ to $(1,i,2,3,\dots,
i-1,i+1,\dots, 2n)$, the products of the factors $(-1)^i$ of (\ref{eq:rec}) is simply $(-1)^{l(\pi)}$
where $l(\pi)$ is the parity of $\pi$. One can therefore rewrite $c(k_1, k_2, \dots, k_{2n})$ as
follows.

\addtocounter{Prop}{-2}

\begin{Prop}[Combinatorial form]  Let the number $2n$ of spinflips be such
that $1\le n\le \frac m2$ and let $1\le k_1 < k_2 < \dots < k_{2n} \le m$ be their
positions. Then
\begin{equation}\label{eq:combi}
c(k_1,k_2,\dots, k_{2n})=c_\uparrow \left(\frac im\right)^n \sum_{q_1,q_2,\dots,q_n\in Q_m}
\sum_{\pi\in \text{\rm Pf}_{2n}} (-1)^{l(\pi)} \prod_{1\le j\le n}d(q_j)
e^{iq_j(k_{\pi(2j-1)}-k_{\pi(2j)})}.
\end{equation}
\end{Prop}

\addtocounter{Prop}{1}

\noindent The probability $\text{Pr}_m(Y_m=n/m)$ is then
\begin{align}
\text{Pr}_m&(Y_m=n/m)\notag\\
& =c_\uparrow\left(\frac im\right)^n \sum_{q_1,q_2,\dots,q_n\in Q_m} \prod_{1\le
\ell\le n}d(q_\ell)
\sum_{1\le k_1<k_2<\dots <k_{2n}\le m}\sum_{\pi\in\text{Pf}_{2n}}(-1)^{l(\pi)}
\prod_{1\le j\le n}e^{iq_j(k_{\pi(2j-1)}-k_{\pi(2j)})}.
\end{align}
Those familiar with classical works on the Ising model will not be surprised to see a pfaffian sum appearing here. (See, for example, \cite{Groeneveld}.)
The continuous limit of (\ref{eq:combi}) can be deduced easily using the
result of the previous paragraph. Using Prop.{} \ref{prop:conti} {\em (ii)}, one finds
that the limit of the probability distribution conditionned to the number of spinflips
is up to a constant
\begin{equation}\label{eq:combiconti}
\sum_{\pi\in{\text{\rm Pf}}_{2n}} \prod_{1\le j\le
n}\sin(\frac12(\theta_{\pi(2j)}-
\theta_{\pi(2j-1)})).
\end{equation}
This expression already exists in the literature, though in a slightly different form.
Burkhardt and Guim \cite{Burk} computed the correlation function $\langle \phi_1(z_1,
\bar z_1)\dots \phi_a(z_a, \bar z_a)\rangle_{\zeta_1, \dots,\zeta_b}$ of fields
$\phi_i, 1\le i\le a$, in the upper-half plane when piecewise constant boundary
conditions are applied along the real axis, with spinflips at $\zeta_j$, $1\le j\le b$. If
the $\phi_i$'s are taken to be the identity, one can argue that $\langle {\mathbf 1}
\rangle_{\zeta_1, \dots,\zeta_b}$ is nothing but the density probability that 
spinflips occur at ${\zeta_1, \dots,\zeta_b}$. With this identification and after
a conformal map of the upper-half plane onto the cylinder (the point at infinity
is deleted), their
expression (16a) is the above expression (\ref{eq:combiconti}). Theirs is obtained from
conformal field theory, ours from the original definition of the Ising model.

It is interesting to remark that only the outer sum on $q_1, q_2, \dots, q_n\in Q_m$ contains the
information about the temperature, through the function $d$. The rest of the expression is
completely combinatorial in nature and rests only upon the introduction of the
anti-commuting
(fermionic) operators
$\xi_q$
of Section \ref{sect:cesaro} in Onsager's solution.
Even though they were introduced
for the square lattice, we will see in Section \ref{sect:TriAndHexa} that similar operators exist for the triangular and hexagonal lattices. It is therefore likely that these operators $\xi_q$ may be introduced for a large
class of two-dimensional lattices and that their commutation relations are independent of the lattice. The function $d$, on the other hand, is likely to depend
on the lattice. It it therefore natural to introduce the function 
$$\widehat{N_n}(q_1, q_2, \dots, q_n)= \sum_{1\le k_1, k_2, \dots, k_{2n}\le m}\sum_{\pi\in\text{Pf}_{2n}}
(-1)^{l(\pi)}
\prod_{1\le j\le n}e^{iq_j(k_{\pi(2j-1)}-k_{\pi(2j)})}.$$
We were not able to find any tractable form for this expression. However note that this expression
is to appear within the sum 
$\sum_{q_j} \prod_{\ell}d(q_\ell)$ where the dependency on all $q$'s is
symmetric. If one decomposes
$\widehat{N_n}$ into its $S_n$-symmetric component, only the fully
symmetric one will contribute to the sum over the $q$'s. Moreover, because the function $d(q)$
is an odd function of $q$, it is sufficient to consider the component in $\widehat{N_n}$ that is 
odd under the exchange of any of the $q$'s. To be more specific let us introduce the following linear
operators. Let $P(q)$ be a polynomial in $e^{iq}$ and $e^{-iq}$. The linear operator
$\text{Odd}_q$ acts as
$(\text{Odd}_q\ P)(q)=\frac12 (P(q)-P(-q))$.
If $P(q_1, q_2, \dots, q_n)$ is a polynomial in $e^{\pm iq_1},e^{\pm
iq_2}, \dots, e^{\pm iq_n}$, the linear operator $\text{Sym}_{q_1,q_2,
\dots,q_n}$ is defined as
$(\text{Sym}_{q_1,q_2,
\dots,q_n}\ P)(q_1,q_2,\dots, q_n)=\frac1{n!}\sum_{\pi\in S_n}
P(q_{\pi(1)},q_{\pi(2)},\dots,q_{\pi(n)})$
where $S_n$ is the permutation group of $n$ elements. In terms of these, the desired probability
is 
\begin{equation}
\text{Pr}_m(Y_m=n/m)=c_\uparrow \left(\frac im\right)^n \sum_{q_1,q_2,\dots, q_n\in Q_m} N_n(q_1, q_2,
\dots, q_n)\prod_{1\le \ell\le n}d(q_\ell)
\end{equation}
where
\begin{align}
N_n&(q_1,q_2, \dots, q_n)\notag\\
&=\left(\prod_{1\le i\le n}{\text{\rm Odd}}_{q_i}\right){\text{\rm 
Sym}}_{q_1,q_2,
\dots,q_n}\sum_{1\le k_1<k_2<\dots<k_{2n}\le m}\ \sum_{\pi\in
\text{\rm Pf}_{2n}} (-1)^{l(\pi)}\prod_{1\le j\le n}e^{i(k_{\pi(2j-1)}-
k_{\pi(2j)})q_j}.
\end{align}
This function $N_n$ is defined for any positive integer $n\le \frac m2$. It is a polynomial
in the $e^{\pm iq_j}$. It is similar in nature to the kernels of Dirichlet and Fej\'er arising
in elementary Fourier analysis. Here $N_n$ is for a discrete Fourier transform of multivariate
functions whose dependency on their variables (momenta) is symmetric and odd.
It is tempting
to call it the Ising kernel. It turns out that this quantity has a very simple expression.
It is shown in the Appendix that $N_n(q_1,q_2, \dots, q_n)$ vanishes whenever there exist $i\neq j$ such that $q_i^2=q_j^2$ and is otherwise
\begin{equation}
N_n(q_1,q_2, \dots, q_n)=
{\displaystyle{\frac1{n!}\left(-\frac{im}2\right)^n\prod_{1\le j\le n}\cot
{\textstyle{\frac12}} q_j}}.
\end{equation}
With this, the summand in $\text{Pr}_m(Y_m)$ becomes, up to a constant, $\prod_{1\le j \le n}
d(q_j) \cot \frac12 q_j$ when non-zero. This function is symmetric and even in all $q$'s. The sum
can be restricted to $q$'s in $Q_m^+=Q_m\cap \mathbb R^+$ by multiplying by $2^n$. And 
because $N_n$ vanishes whenever some of the $q$'s coincide up to sign, the probability corresponds to equation \eqref{eq:qm}.

\begin{figure}
\begin{center}\leavevmode
\includegraphics[bb = 65 100 750 580,clip,width=0.8\textwidth]{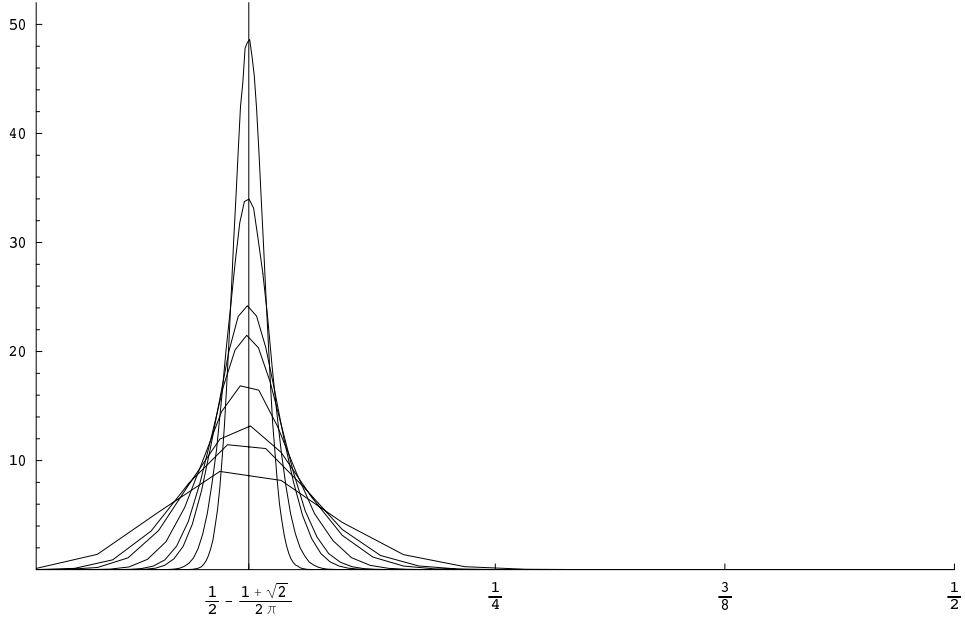}
\end{center}
\caption{Probability distributions of the variables $Y_m$ for
several values of $m$.\label{fig:spinflip}}
\end{figure}

After long calculations, an agreement with explicit simulations may be sought for reassurance.
The Figure \ref{fig:spinflip} was drawn for that purpose. The cases $m=30, 48, 60,
99,\allowbreak 157,200,\allowbreak 397$ and $800$ are plotted. They are the results of Monte-Carlo simulations except
for the two smallest lattices that represent exact calculations. In the computer experiments
the length of the cylinder was twice as long as the circumference and the smallest sample ($5\times10^5$)
was for the $m=800$ cylinder. As an example of measurements of $\sigma$, this latter case gave
$\hat\sigma=0.05372$. The reader will notice both
even and odd $m$'s here even though the proposition was proved for $m$ even only.

The last result of this section is not a consequence of Proposition \ref{prop limit spinflips} as such but
it does follow from the technique used in its proof.

\begin{Cor} The probability $c_\uparrow(m)$ of having only spins $+$ at the boundary of a
half-infinite cylinder (at the critical temperature) behaves as
\begin{equation}\label{eq:cfleche}
c_\uparrow(m)\approx \frac1{\sqrt2}\left(2(\sqrt2-1)e^{2G/\pi}\right)^{-\frac m2}e^{-\frac{\pi}{24m}}
\end{equation}
for $m$ large enough. $G$ is Catalan's constant: $G=\sum_{k=0}^\infty(-1)^k(2k+1)^{-2}$.
\end{Cor}

\noindent{\scshape Proof: }  The probability $c_\uparrow$ is half the probability of having no spinflip
and, by the preceding proof,
$c_\uparrow(m)={\textstyle{\frac12}}\text{Pr}(n=0)=(2\prod_{1\le i\le M}(1+f_{M,i}))^{-1}$
with $M=\frac m2$ and $f_{M,i}=d(q_i)\cot\frac12 q_i$, $q_i=(2i-1)\pi/m\in Q_m^+$. A direct
calculation leads to
$$1+f(q)=\frac1{\sin \frac q2}\frac{2\sqrt2 -1-\cos q+\sqrt{(1-\cos q)(3-\cos q)}}{2\sqrt{1-\cos q}+\sqrt2\sqrt{3-\cos q}}.$$
Using standard tables (see for example paragraph 6.1.2 of \cite{PBM}), one finds
$\prod_{j=1}^{\frac m2}\sin \frac{q_j}2=\frac1{2^{(m-1)/2}}$.
Therefore 
\begin{align*}
\prod_{1\le i\le M}(1+f_{M,i}) &=\exp \sum_{1\le i\le M}\log (\sin q_i/2+d(q_i)\cos q_i/2)-\log(\sin(q_i/2))\\
&=2^{(m-1)/2} \exp \sum_{q\in Q^+_m} \log h(q)
\end{align*}
where
$$h(q)=\frac{2\sqrt2-1+\cos q+\sqrt{(1+\cos q)(3+\cos q)}}{2\sqrt{1+\cos q}+\sqrt2\sqrt{3+\cos q}}.$$
This function $h$ is the remaining factor in $(1+f)$ evaluated at $\pi-q$. Note that the set $Q^+_m$ is
stable under the operation $q\rightarrow \pi-q$. Moreover $h$ has the following useful properties: it is
analytic on $(-\pi,\pi)$, even and takes the value $1$ at $q=0$. One can write
$\log h(q)=\sum_{i\ge2, \text{even}}a_i q^i$
and therefore
$\sum_{q\in Q^+_m} \log h(q) = \sum_{i\ge2, \text{even}} a_i \sum_{j=1}^{m/2}q^i_j$.
The leading terms (in $m$) in the inner sum are
$\sum_{j=1}^{m/2} q^i_j=\pi^i\left(\frac m{2(i+1)}-\frac 1{12}\frac im + \mathcal{O}\left(\frac
1{m^2}\right)\right)$. The first term can be resummed as follows
$$\sum_{i\ge2, \text{even}} a_i \frac{\pi^i m}{2(i+1)}=\frac m{2\pi}\sum_{i\ge2, \text{even}}\int_0^\pi a_i x^i dx=
\frac m{2\pi}\int_0^\pi \log h(q) dq$$
and the second as
$$-\frac 1{12}\sum_{i\ge2, \text{even}} a_i \frac{\pi^i i}m= - \frac{\pi}{12m}\left. \frac{d\ }{dq}\log h(q)\right|_{q=\pi^-}
=\frac{\pi}{24m}.$$
The integral appearing in the first term is again somewhat difficult. Rewriting the integrand as follows and using {\em Mathematica}, we were able to obtain
\begin{equation*}
\int_0^\pi \log h(q) dq = \int_0^\pi \log \left( (1-\frac1{\sqrt2})(\sqrt{1-\cos q}+\sqrt{3-\cos q})\right) dq 
= \pi \log (\sqrt2 -1) + 2G
\end{equation*}
where $G$ is Catalan's constant.
Gathering the various terms, we get \eqref{eq:cfleche}. \hfill$\Box$

%
%
%

\section{Behavior at the boundary for the 
triangular and hexagonal lattices}\label{sect:TriAndHexa}
\begin{figure}
\begin{center}\leavevmode
\includegraphics[width = 10cm]{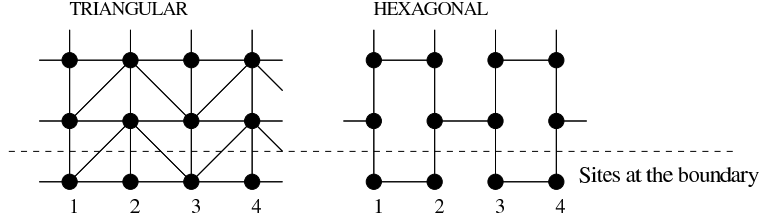}
\end{center}
\caption{Triangular and hexagonal lattices with our particular choice
  of boundaries.
\label{fig:lattice-transfer}}
\end{figure}

In this section we extend the techniques used for the square lattice to compute the quantity $c^s(k_1,k_2)$ for the hexagonal and
the triangular cases. These lattices are characterized by a
connectivity of 6 and 3 respectively except at the
boundary, where more than one choices can be made. 
Our choice is drawn in figure \ref{fig:lattice-transfer}. 
The square-shaped representation will be helpful for the computations of the transfer matrix that follows.
 
The main result of the section asserts that
the thermodynamic limit of $c^s(k_1,k_2)$ at critical temperature behaves
as in the square lattice case. This yields
some evidence of the universality of the critical behavior at the boundary. 
\begin{Prop}
Let $m$ be even, $k$ be such that $1\le k\le 2m$ and set
$\theta=\frac{2\pi k}{2m}$. Let $p,n$ and $r$ as in Lemma \ref{prop:lemma2}
with $\gamma=4$. Denote by $\lim$ the process of taking the limit $k,
m\rightarrow \infty$ while keeping both $\theta$ and $r$ fixed. Then at criticality for $k= k_2-k_1$
$$\lim \frac{c^s(k_1,k_2)}{c_{\uparrow}} =\frac{C_\text{\rm lattice}}{\sin\theta/2}
$$
where $C_\text{\rm lattice}$ is either $C_\text{\rm tri}=2(2-\sqrt{3})$ or $C_\text{\rm hex}=4/3$ if $k$ is even and $2(\sqrt{3}-1)/3$ if $k$ is odd.
\label{prop hextri}
\end{Prop}
Even though we focused on the case of two spinflips, it is possible to generalize inductively our result 
to $c^s(k_1,...,k_n)$ in the same way as in Section 2. The result would be similar to the square lattice with 
the appropriate constant $C_{\text{lattice}}$.

The proposition implies that, for the hexagonal lattice,
various limits can be obtained for
the same $\theta$. To understand this, one should look at Figure \ref{fig:lattice-transfer}. 
For the triangular lattice, a spinflip on an even site is actually equivalent (up to a reflection) to a spinflip on an odd site. This is not true for the hexagonal lattice.  The case of a spinflip on an odd site is different from a spinflip on an even site: in the latter the sites are not nearest-neighbors whereas in the first, they are. This asymmetry persists in the continuum limit. 
A parallel with
quantum field theory is useful.
It is impossible
physically to measure this probability
precisely for a given $\theta$.
A measurement that {\em is} possible
is to obtain the probability that the
two spinflips occur within a distance $\theta\in [ \theta_{\text{min}},\theta_{\text{max}}]$. To calculate this probability (or this ``smeared correlation function'') from
the above result, one would have to use the average of
$C_\text{\rm hex}^{\text{\rm even}}$ and
$C_\text{\rm hex}^{\text{\rm odd}}$ as the sites with $k$
even and those with $k$ odd have the same density. (Of course
this above limiting distribution is not 
normalizable and a cut-off becomes
necessary.)

We start by reviewing the
diagonalization of the transfer matrix in Section \ref{sect:TriAndHexapart1} following the work of Houtappel \cite{Hout}. 
Then we construct in Section \ref{sect:TriAndHexapart2} an equation similar to \eqref{eq:omega} 
for the eigenvector with the largest
eigenvalue. We get, in a similar way as
in Section \ref{sect:cesaro}, an expression for $c^s(k_1,k_2)$ which
involves functions that play the same role as $d(\nu,q)$ in the square
case. These functions are investigated only at critical
temperature. As before the critical behavior is related to the jump of these functions at $q=0$.\

\subsection{Transfer matrix for the triangular and hexagonal lattices}
\label{sect:TriAndHexapart1}
The element of the transfer matrices for our particular choice of
lattices and boundaries for a transfer from row $\sigma$ to row
$\sigma'$ with $2m$ sites ($m$ even), i.e.\ the difference of Gibbs weights between
these two rows, is
\begin{equation*}
\begin{aligned}
T_{\tri}(\sigma',\sigma)&=\exp\left(\nu\left\{\sum_{j=1}^{2m}
    \sigma_j\sigma'_j+\sum_{j=1}^{2m} \sigma'_j\sigma'_{j+1}+\sum_{j=1}^m(\sigma'_{2j-1}+\sigma'_{2j+1})\sigma_{2j}\right\}\right)\\
T_{\hex}(\sigma',\sigma)&=\sum_{\sigma^I}\exp\left(\nu\left\{\sum_{j=1}^{2m}
    (\sigma_j+\sigma'_j)\sigma^I_j+\sum_{j=1}^m(\sigma'_{2j-1}\sigma'_{2j}+\sigma^I_{2j}\sigma^I_{2j+1})\right\}\right)
\end{aligned}
\end{equation*}
where $\sigma_{2m+1}\equiv \sigma_1$ and the sum over $\sigma^I$ is
the sum over all possible configurations of the intermediate row. It is
convenient to write the previous expressions in terms of matrices that
contain interactions along columns only or along rows only to follow
thereafter the calculation for the square lattice. We explain here the
triangular case. The hexagonal case is done exactly the same way and
only major steps are given.

\begin{figure}
\begin{center}\leavevmode
\includegraphics[width = 13cm]{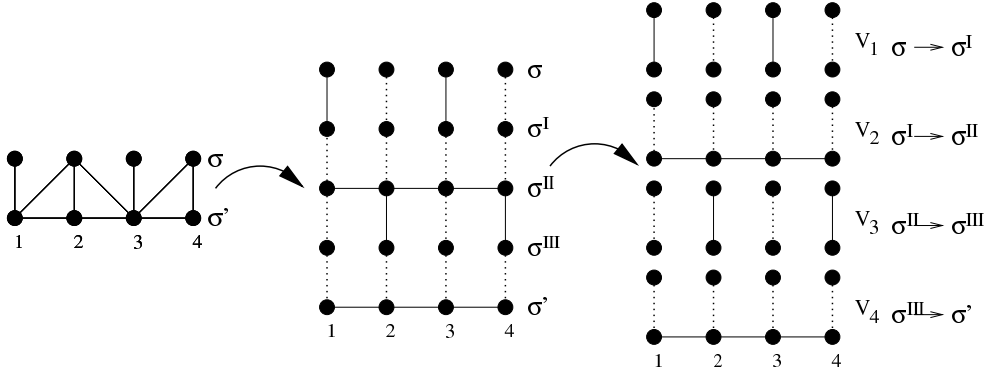}
\end{center}
\caption{Decomposition of the transfer
  matrix into $V_1, V_2, V_3, V_4$
  for the triangular case.
\label{fig:lattice-transfer2}}
\end{figure}

We decompose the transfer matrix from $\sigma$ to $\sigma'$ into
the product of several transfer matrices.
Figure \ref{fig:lattice-transfer2} depicts this process. Dotted bonds
represent the identification of two sites (in other words a weight of
$1$ for the initial state and of $0$
for the other state). Plain lines are the usual ferromagnetic
bonds. With this particular choice of decomposition
interactions along rows and along columns lie in different matrices.
One gets the following
representation of the transfer matrix by summing over
all possible intermediate configurations 
\begin{equation}
T_{\tri}(\sigma',\sigma)=\sum_{\sigma^I,\sigma^{II},\sigma^{III}}V_4(\sigma',\sigma^{III})V_3(\sigma^{III},\sigma^{II})V_2(\sigma^{II},\sigma^I)V_1(\sigma^I,\sigma)
\end{equation}
where the 
$V_i$'s are defined as follows 
\begin{equation}
\begin{aligned}
V_1(\sigma',\sigma)&=\exp\left(\nu\sum_{j \text{ odd}}
  \sigma_j\sigma'_j\right)\prod_{j\text{ even}}\delta_{\sigma_j,\sigma'_j} 
  \hspace{0.4cm}
   V_3(\sigma',\sigma)=\exp\left(\nu\sum_{j \text{ even}}
  \sigma_j\sigma'_j\right)\prod_{j\text{ odd}}\delta_{\sigma_j,\sigma'_j}\\
V_2(\sigma',\sigma)&=V_4(\sigma',\sigma)=\exp\left(\nu\sum_{j}
  \sigma'_j\sigma'_{j+1}\right)\prod_{j}\delta_{\sigma_j,\sigma'_j}.
\end{aligned}
\end{equation}
 
Second we rewrite these matrices in terms of the operators
$\tau^i_k$, $k=1,...,2m$ and $i=1,2,3$, and of
the operators $\pi_k$, $\rho_k$ and $P$ 
(all defined in Section \ref{sect:cesaro} and
in \cite{thompson})
\begin{equation}
\begin{aligned}
V_1&=(2\sinh 2\nu)^{m/2}\exp\left(i\nu^*\sum_{j \text{ odd}}
  \pi_{j}\rho_{j}\right)
     \hspace{0.4cm}
V_3=(2\sinh 2\nu)^{m/2}\exp\left(i\nu^*\sum_{j \text{ even}}
  \pi_{j}\rho_{j}\right)\\
V_2&=V_4=\exp(i\nu\pi_1\rho_{2m} P)\exp\left(-i\nu\sum_{j=1}^{2m-1}\pi_{j+1}\rho_j\right)
\end{aligned}
\end{equation}
where $\sinh\nu\sinh\nu^*=1$. As in the square case, the parity
operator $P$ must commute with each $V_i$ because
the Gibbs weights are invariant under $\sigma\rightarrow -\sigma$. It is
therefore possible to diagonalize simultaneously $T$ and $P$. 
Hence we write
$T=\frac12 (1+P)T^+ + \frac12 (1-P)T^-$ where $T^\pm$ 
acts in a non-trivial way on the even/odd subspace 
and has eigenvalue zero for
odd/even eigenvectors. 
Taking this into account in the previous
expression of $V_2$ and replacing $\pi_k$ and $\rho_k$ by their
expression in terms of $a_k$, we get
\begin{equation}
\begin{aligned}
V_1&=(2\sinh 2\nu)^{m/2}\exp\left(-2\nu^*\sum_{j \text{ odd}}
  a_j^\dagger a_j -\frac{1}{2}\right)
  \hspace{0.2cm}
V_3=(2\sinh 2\nu)^{m/2}\exp\left(-2\nu^*\sum_{j \text{ even}}
  a_j^\dagger a_j -\frac{1}{2}\right)\\
V_2^\pm&=V_4^\pm=\exp\left(\nu\sum_{j}(a_{j+1}^\dagger
  -a_{j+1})(a_{j}^\dagger+a_{j})\right)
\label{eqn:transfer tri}
\end{aligned}
\end{equation} 
with $T^+_{\tri}=V^+_4V_3V^+_2V_1$ and similarly for $T^-_{\tri}$. The
condition of parity imposes $a_{2m+1}=\mp a_1$ for $V^\pm$. 

We claim that the eigenvector $\omega$ with highest eigenvalue lies in the even subspace. 
Indeed the operator $P$ flips a configuration $\sigma$ into $-\sigma$ and
it acts as $-I$ in the odd sector. If $\omega$ lies in the odd sector, then $P\omega=-\omega$ and its coefficients 
in the basis of configurations must satisfy: $c_{\sigma_1\sigma_2\dots \sigma_{2m}}
=-c_{-\sigma_1-\sigma_2\dots -\sigma_{2m}}$. It follows that some coefficients are negative
or they are all zero, contradicting Frobenius theorem. We shall thus be interested in diagonalizing $T^+$. 

The particular form of $V_i$ for $i=2,4$ leads us to
define two Fourier transforms:
\begin{equation}
\begin{aligned}
a_k&=
\begin{cases} \frac{e^{i\pi/4}}{\sqrt{m}} \sum_{q\in Q_m} e^{ikq}\eta_q, &
  \text{  for $k$ odd}\\
\frac{e^{i\pi/4}}{\sqrt{m}} \sum_{q\in Q_m} e^{ikq}\delta_q, &
  \text{  for $k$ even}\\
\end{cases}\\
{\text{\rm where\ \ }} Q_m&=\left\{\frac{(2l-1)\pi}{2m}:-m/2+1\leq l\leq m/2\right\}.
\label{eqn:fourier}
\end{aligned}
\end{equation}
Again, $m$ is assumed even. The particular definition of $Q_m$
allows us to invert the previous relations to get 
\begin{equation}
\begin{aligned}
\eta_q&=\frac{e^{-i\pi/4}}{\sqrt{m}}\sum_{k \text{ odd}}e^{-iqk}a_k,\qquad \quad
\delta_q&=\frac{e^{-i\pi/4}}{\sqrt{m}}\sum_{k \text{
    even}}e^{-iqk}a_k.
\end{aligned}
\end{equation} 
The anti-commutation relations of the $a_k$ induce relations for the Fourier transforms
\begin{equation}
\begin{aligned}
\{\eta_q,\eta_{q'}\}=\{\delta_q,\delta_{q'}\}=0, \quad
\{\eta_q,\delta_{q'}\}=\{\eta_q,\delta_{q'}^{\dagger}\}&=0,\\
\{\eta_q,\eta_{q}^{\dagger}\}=\{\delta_q,\delta_{q}^{\dagger}\}=1,
\quad\{\eta_q,\eta_{q'}^{\dagger}\}=\{\delta_q,\delta_{q'}^{\dagger}\}&=0 \quad\text{if $q\neq q'$}.
\end{aligned}
\label{eqn:relationFourier}
\end{equation}

These relations decouple the various sectors of momentum $q$, $-\pi/2<q<\pi/2$. 
As we shall now see the $V$'s mix the sector of a given
momentum $q$ with that of momentum $-q$, but with no other.
Therefore it is possible to express $T$ in terms of a
tensor product of operators acting on specific sectors of momenta $q$ and $-q$.
A direct substitution of (\ref{eqn:fourier}) in (\ref{eqn:transfer tri}) gives the precise form of these
operators. 
We get $T^+_\text{tri} =(2\sinh
2\nu)^{m/2}\prod_{0<q<\pi/2}V_q$ for
$V_q=V^{+}_{4,q}V_{3,q}V^{+}_{2,q}V_{1,q}$ where
\begin{equation}
\begin{aligned}
V_{1,q}&=\exp\left(-2\nu^*\left(\eta_q^{\dagger}\eta_q+\eta_{-q}^{\dagger}\eta_{-q}-1\right)\right)
\hspace{0.4cm}
V_{3,q}=\exp\left(-2\nu^*\left(\delta_q^{\dagger}\delta_q+\delta_{-q}^{\dagger}\delta_{-q}-1\right)\right)\\
V^+_{2,q}=V^+_{4,q}&=\exp\left(\nu e^{-iq}\left(-i
    (\eta_q^{\dagger}\delta_{-q}^{\dagger}+\eta_{-q}\delta_q+\delta_{q}^{\dagger}\eta_{-q}^{\dagger}+\delta_{-q}\eta_q)\right.\right.\\
    &\phantom{\exp\big(\nu e^{-iq}\big( -i}
    +\left.\left. (\delta_{-q}^{\dagger}\eta_{-q}+\eta_{q}^{\dagger}\delta_q+\delta_q^{\dagger}\eta_{q}+\eta_{-q}^{\dagger}\delta_{-q})\right)+\star\right)
\end{aligned}
\end{equation}
where $\star$ holds for a term identical to the preceding one but
with $q\rightarrow -q$. 
A similar calculation of the transfer matrix for the hexagonal case yields
\begin{equation}
\begin{aligned}
W_{1,q}=W_{3,q}&=\exp\left(-2\nu^*\left((\eta_q^{\dagger}\eta_q+\delta_q^{\dagger}\delta_q-1)+\star\right)\right)\\
W^+_{2,q}&=\exp\left(\nu
  e^{-iq}\left(-i(\eta_q^{\dagger}\delta_{-q}^{\dagger}+\eta_{-q}\delta_q)+(\eta_{q}^{\dagger}\delta_q+\delta_{-q}^{\dagger}\eta_{-q})\right)+\star\right)\\
W_{4,q}&=\exp\left(\nu e^{-iq}\left(-i(\delta_{q}^{\dagger}\eta_{-q}^{\dagger}+\delta_{-q}\eta_q)+(\delta_q^{\dagger}\eta_{q}+\eta_{-q}^{\dagger}\delta_{-q})\right)+\star\right)
\end{aligned}
\end{equation}
for $T^+_\text{hex}=(2 \sinh 2\nu)^m\prod_{0<q<\pi/2}W_q$ for $W_q=W_{4,q}W_{3,q}W^+_{2,q}W_{1,q}$.

Recall that each $V_q$ and $W_q$ appearing in the tensor product are operators acting on a space labeled by 
a pair $(q,-q)$ with $q\in Q_m$.
Each space (there are $m/2$ in total) is of dimension $2^4$.
A basis for a given space labeled by 
$q$ is constructed by applying the fermionic operators
$\eta_{\pm q}$, $\delta_{\pm q}$ and their conjugates on the
vacuum denoted $|0\rangle$. (It should be stressed that $\vac$ {\em is not} an
eigenvector of $V_q$. See \cite{SML}.)
One can then recognize five
subspaces of a space $q$ that are invariant under the action of $V_q$ and $W_q$. 
They are spanned by: 
\begin{equation}
\begin{aligned}
&\{\delta^\dagger_{q}
\eta^\dagger_{q}\vac\}\\
&\{\eta^\dagger_{q}\vac, \delta^\dagger_{q}\vac, \delta^\dagger_{-q}
\delta^\dagger_{q}\eta^\dagger_{q}\vac, \delta^\dagger_{q}
\eta^\dagger_{-q}\eta^\dagger_{q}\vac\},\\
&\{\vac,\delta^\dagger_{-q}\eta^\dagger_{q}\vac,\delta^\dagger_{q}\eta^\dagger_{-q}\vac
,\delta^\dagger_{-q}\delta^\dagger_{q}\vac,\eta^\dagger_{-q}\eta^\dagger_{q}\vac,
\delta^\dagger_{-q}\delta^\dagger_{q}\eta^\dagger_{-q}\eta^\dagger_{q}\vac\},\\
&\{\eta^\dagger_{-q}\vac, \delta^\dagger_{-q}\vac, \delta^\dagger_{-q}
\delta^\dagger_{q}\eta^\dagger_{-q}\vac, \delta^\dagger_{-q}
\eta^\dagger_{-q}\eta^\dagger_{q}\vac\},\\
&\{\delta^\dagger_{-q}\eta^\dagger_{-q}\vac\}. 
\end{aligned}\label{eq:baseDes16}
\end{equation}
We will refer to these subspaces
respectively as subspaces of momentum $2q$, $q$, $0$, $-q$ and $-2q$.
Because the highest eigenvalue of the transfer matrix $T$ is 
not degenerate,
none of the highest eigenvalue of the $V_q$'s can be either.
Hence, because the actions of $V_q$ on the subspaces of momenta $q$ and
$-q$ are identical (similarly for $2q$ and $-2q$), 
$\omega$ must lie in
the subspace of momentum $0$. 
To get $\omega$, it thus remains to diagonalize the block acting on the space of momentum $0$, which is of dimension $6$. 
This can be done by writing the representation of
$V_q$ in this subspace in the basis \eqref{eq:baseDes16}. Symbolic manipulation programs
can be used to take exponentials and to perform the diagonalization. We will not write these expressions here
but they will be used in the next section to compute $c^s(k_1,k_2)$.

\subsection{Computation of $c^s(k_1,k_2)$}
\label{sect:TriAndHexapart2}
The first step in computing $c^s(k_1,k_2)$ is to construct two operators that annihilate
$\omega$, instead of one for the square lattice. It then remains to apply them to $\omega$ in the basis of
the spinflips to get $c^s(k_1,k_2)$. We will need three
important facts. The results apply to both the hexagonal and
triangular cases. $V_q$ stands here for either $V_q$ or $W_q$.
\begin{Lem}
For all $q>0$ in $Q_m$, the operator $V_q$ satisfies the
relation $\phi_q^TV_q^T\phi_qV_q=1$
where
$\phi_q=\left(\delta_q^\dagger+\delta_q\right)(\delta_{-q}^\dagger+\delta_{-q})\left(\eta_q^\dagger+\eta_q\right)(\eta_{-q}^\dagger+\eta_{-q})$
and ``$\ ^T$'' denotes the transposed operator.
\label{lem:identity}
\end{Lem}

\noindent{\scshape Proof:} By definition of $\delta_q$ and $\eta_q$ in
terms of Pauli matrices, one gets
$\phi_q^T=\phi_q$ and $\phi_q^2=1$. Hence it suffices to show that
$\phi_q^TV_{i,q}^T\phi_qV_{i,q}=1$, $i=1,2,3,4$. This is
directly verified by expressing every operator in terms of $\delta_q$ and
$\eta_q$ and by using their anti-commutation relations (\ref{eqn:relationFourier}).
\hfill $\Box$

\begin{Prop}
Let $\omega_1$ and $\omega_2$ be two eigenvectors of $V_q$ with
eigenvalues $\lambda_1$ and $\lambda_2$ respectively. Then either
$\lambda_1=1/\lambda_2$ or $(\omega_1,\omega_2)_{\phi_q}=0$, where
$(\ \cdot\ ,\ \cdot\ )_{\phi_q}$ is the bilinear form induced by $\phi_q$: $(u,v)_{\phi_q}
\equiv u^T\phi_qv$.
\end{Prop}

\noindent{\scshape Proof:} By direct calculations, using Lemma
\ref{lem:identity}.\hfill $\Box$

\begin{Cor} If $v_q$ is an eigenvector of $V_q$ with eigenvalue $\lambda\neq 1$,
then $(v_q,v_q)_{\phi_q} =0$.
\label{prop:corollary8}
\end{Cor}
 
We note that the subspace of momentum $0$ is an invariant subspace of
$\phi_q$ so that the corollary holds on this subspace. 
If $\omega_q$ denotes the component of $\omega$ in the subspace $q$ of momentum $0$, we must have by the corollary that
$(\omega_q,\omega_q)_{\phi_q} =0$, since the action of $V_q$ and $W_q$ on the subspace is different than the identity.
In the basis of \eqref{eq:baseDes16}, this becomes for $\omega_q=(v_1,v_2,v_3,v_4,v_5,v_6)$
\begin{equation}
v_1v_6-v_2v_3-v_4v_5=0.
\label{eq:identity}
\end{equation}
We stress that all the above statements hold at any temperature.

The desired operators are now constructed. By analogy with the square
case we choose the form
\begin{equation}
\xi_q=a\eta_q+b\delta_q+c\eta^\dagger_{-q}+d\delta^\dagger_{-q}.
\label{eq:xi tri}
\end{equation}
where $a$, $b$, $c$, $d$ may depend on $q$.
We argued previously that $\omega$ belongs to the
subspace of momentum $0$. Hence, from the definition of $\xi_q$, $\xi_q\omega_q$ must lie in the subspace of momentum $-q$. 
The equation $\xi_q\omega_q=0$ written in the basis \eqref{eq:baseDes16} becomes in this $4$-dimensional subspace
\begin{equation*}
\begin{pmatrix}
v_5 & -v_3 & -v_1 & 0 \\
-v_2 & -v_4 & 0 & v_1 \\
-v_6 & 0 & v_4 & v_3 \\
0 & v_6 & v_2 & -v_5 \\
\end{pmatrix}\begin{pmatrix} a\\b\\c\\d\end{pmatrix}=0.
\end{equation*}
If we assume that the components $v_i$ of $\omega$ are non-zero,
then the $4\times4$ matrix above can be shown to be of rank $2$ by (\ref{eq:identity}).
It is easily checked that the kernel is spanned by $(-v_2,v_5,0,v_6)$ and
$(v_4,-v_2,v_6,0)$. These vectors yield the two desired operators that we now refer to as
$\xi_q$ and $\xi'_q$ respectively.  

The coefficient $c^s(k_1, k_2)$ can now be calculated as in section
\ref{sect:cesaro}. We first write $\omega$ in the basis of the
spinflips:
$\omega=c_\uparrow\sigma_\uparrow+c_\downarrow\sigma_\downarrow+\sum_{l=1}^n\sum_{1\leq
  k_1<...<k_{2l}\leq 2m}c^\pm(k_1,...,k_{2l})(\pm;k_1,...,k_{2l})$.
We compute the action of $\xi_q$ and $\xi'_q$,
$0<q<\pi/2$, on $\omega$ in the space of momentum $0$
taking $\xi_q$ and $\xi_q'$ in the form (\ref{eq:xi tri}). We recall
that the action of these operators is well understood in the basis of
spinflips if one writes $\xi_q$ and $\xi'_q$
in terms of operator $\pi_k$ and
$\rho_k$ that change the sign of the spins from the site $1$ to $k-1$
and from $1$ to $k$ respectively. The result of the action must be $0$. 
We calculate $c^\pm(1,k)$ and $c^\pm(k_1,k_2)$ follows by translational invariance.
The projections of $\xi_q\omega$ and $\xi'_q\omega$ on $\sigma_\uparrow$ yield two equations:
\begin{equation}
\begin{aligned}
c_\uparrow\left(-v_3e^{-iq}-v_5-iv_6\right)=&\sum_{k
  \text{ even}}c^-(1,k)e^{-iqk}\left(-v_3e^{iq}+v_5-iv_6\right)\\
+&\sum_{k \text{ odd}} c^-(1,k)e^{-iqk}\left(v_3-v_5e^{iq}-iv_6e^{iq}\right)\\
c_\uparrow\left(-v_4e^{-iq}+v_2+iv_6e^{-iq}\right)=&\sum_{k
  \text{ even}}c^-(1,k)e^{-iqk}\left(-v_4e^{iq}-v_2-iv_6e^{-iq}\right)\\
+&\sum_{k \text{ odd}} c^-(1,k)e^{-iqk}\left(v_4+v_2e^{iq}-iv_6\right).
\end{aligned}
\end{equation}

These two equations can be solved for $\sum_{k
\text{ odd}}c^-(1,k)$ and $\sum_{k \text{ even}}c^-(1,k)$.
The inverse Fourier transform on $q$ yields $c^-(1,k)$. 
By translational invariance, we get for any  $k_1$, $k_2$:
\begin{equation}
c^s(k_1,k_2)=\begin{cases} \frac{-i}{2m}\sum_{q\in
    Q_m}e^{iq(k_2-k_1)}d_{\text{even}}(\nu,q), \quad &\text{if $k_2-k_1+1$ is
    even,}\\
 \frac{-i}{2m}\sum_{q\in
    Q_m}e^{iq(k_2-k_1)}d_{\text{odd}}(\nu,q), \quad &\text{if $k_2-k_1+1$ is
    odd,}
 \end{cases}
\end{equation}
where 
\begin{equation*}
\begin{aligned}
d_{\text{even}}(\nu,q)&=\frac{-2(v_1+v_6)+2i(v_5-v_4)}{\sin q (v_6-v_1)-\cos q(v_4+v_5)+(v_3-v_2)},\\
d_{\text{odd}}(\nu,q)&=\frac{2\cos q (1-v_1)+2\sin q (v_4+v_5)+2i(v_2+v_3)}{\sin q (v_6-v_1)-\cos q(v_4+v_5)+(v_3-v_2)}.
\end{aligned}
\end{equation*}
The components $v_1,...,v_6$ of $\omega_q$ are obtained following Section \ref{sect:TriAndHexapart1}.
The expressions for a generic temperature $\nu$ are rather heavy. 
We write down the result for $\nu=\nu_c$. The use of symbolic manipulation programs was essential to obtain
the expressions in these relatively simple forms. 

\smallskip

\noindent{\scshape Triangular Lattice}
\begin{equation}
\begin{aligned}
d^\text{tri}_{\text{even}}(\nu_c,q)&=
  \begin{cases}\frac{2\left(-3 - \sin
    q + 2\left(\cos\frac{q}{2}+\sin\frac{q}{2}\right)\sqrt{3+\sin
    q}\right)+2i\left(2\cos{q} - \left(\cos\frac{q}{2}-\sin\frac{q}{2}\right)\sqrt{3+\sin
    q}\right)}{\sqrt{3}(1+3\sin q)},
    & \text{ if $0\leq q\leq\frac{\pi}2$};\\
-\text{Re}\left[d^\text{tri}_{\text{even}}(\nu_c,-q)\right]
+i\ \text{Im}\left[d^\text{tri}_{\text{even}}(\nu_c,-q)\right], &\text{ if $-\frac{\pi}2\leq q<0$};\end{cases}\\
d^\text{tri}_{\text{odd}}(\nu_c,q)&=\begin{cases}\frac{4\left(\cos\frac{q}{2}-\sin\frac{q}{2}\right)\left(\cos\frac{q}{2}+\sin\frac{q}{2}-\frac{1}{2}\sqrt{3+\sin
    q}\right)}{1+3\sin q},
    &\text{ if $0\leq q\leq\frac{\pi}2$};\\
-d^\text{tri}_{\text{odd}}(\nu_c,-q), &\text{ if $-\frac{\pi}2\leq q<0$}.
\end{cases}
\end{aligned}
\end{equation}

\smallskip

\noindent{\scshape Hexagonal Lattice}
\begin{equation}
\begin{aligned}
d^\text{hex}_{\text{even}}(\nu_c,q)&=\begin{cases}
-\frac{2\left(\sqrt{3}-1\right)\left(3+\sin q
    -\sqrt{3}\left(\cos\frac{q}{2}+\sin\frac{q}{2}\right)
    \sqrt{3+\sin q}\right)}{\sin{q}\left(3+\sin{q}\right)},
 \quad &\text{ if $0\leq q\leq\frac{\pi}2$};\\
-d^\text{hex}_{\text{even}}(\nu_c,-q),\quad  &\text{ if $-\frac{\pi}2\leq q<0$};
\end{cases}\\
d^\text{hex}_{\text{odd}}(\nu_c,q)&=\begin{cases}
\frac{\left(\cos\frac{q}{2}-\sin\frac{q}{2}\right)\left(-2\sqrt{3}\sqrt{3+\sin q}
  +2\left(\cos\frac{q}{2}+\sin\frac{q}{2}\right)
  \left(3+\sin{q}\right)\right)}{\sin{q}\left(3+\sin{q}\right)},
&\text{ if $0\leq q\leq\frac{\pi}2$};\\
-d^\text{hex}_{\text{odd}}(\nu_c,-q), &\text{ if $-\frac{\pi}2\leq q<0$}.
\end{cases}
\end{aligned}
\end{equation}

\smallskip

We gather in a lemma the properties of the above functions that are relevant for the limit $m\to\infty$. 
Its proof is similar to that of Lemma \ref{prop:lemma1}.
The continuum limit of $2mc^s(k_1,k_2)/c_{\uparrow}$ in Proposition \ref{prop hextri} follows directly from Lemmas \ref{prop:lemma2} and \ref{lem:properties tri-hex}, with $r\theta=2\pi k/\gamma=\pi k/2$.
\begin{Lem}\label{prop:lemma9}
The four functions 
$d^{\text{\rm tri}}_{\text{\rm odd}}$,
$d^{\text{\rm tri}}_{\text{\rm even}}$,
$d^{\text{\rm hex}}_{\text{\rm odd}}$,
$d^{\text{\rm hex}}_{\text{\rm even}}$ on $[-\frac{\pi}2,\frac{\pi}2]$
have the following properties:

\noindent {\em (i)} their restriction to $[0,\frac{\pi}2]$ satisfy
the requirements put on $z$ in Lemma \ref{prop:lemma2};

\noindent {\em (ii)} their parity and some of their relevant values are as follows.
{\rm
\begin{center}
\begin{tabular}{|c|c|c|c|c|c|}
\hline
  & values in & & parity & $(\nu_c,0^+)$ & $(\nu_c,\frac{\pi}2)$ \\
\hline
\hline
$d^{\text{\rm tri}}_{\text{\rm odd}}$ & $\mathbb R$ & & odd & $2(2-\sqrt3)$ & 0 \\
\hline
\multirow{2}{0.7cm}{$d^{\text{\rm tri}}_{\text{\rm even}}$}&
\multirow{2}{0.36cm}{$\mathbb C$} & 
$\text{\rm Re\ } d^{\text{\rm tri}}_{\text{\rm even}}$ & 
odd & 
$2(2-\sqrt3)$ & \\ 
& & $\text{\rm Im\ } d^{\text{\rm tri}}_{\text{\rm even}}$ & even & 
$2(2-\sqrt3)$ & 0\\
\hline
\hline
$d^{\text{\rm hex}}_{\text{\rm odd}}$ & $\mathbb R$ & & odd & $\frac43$ & 0 \\
$d^{\text{\rm hex}}_{\text{\rm even}}$ & $\mathbb R$ & & odd & $\frac23(\sqrt3-1)$ &  \\
\hline
\end{tabular}
\end{center}
}
\label{lem:properties tri-hex}
\end{Lem}

\begin{figure}
\begin{center}\leavevmode
\includegraphics[bb = 50 320 580 700,clip,width = 13cm]{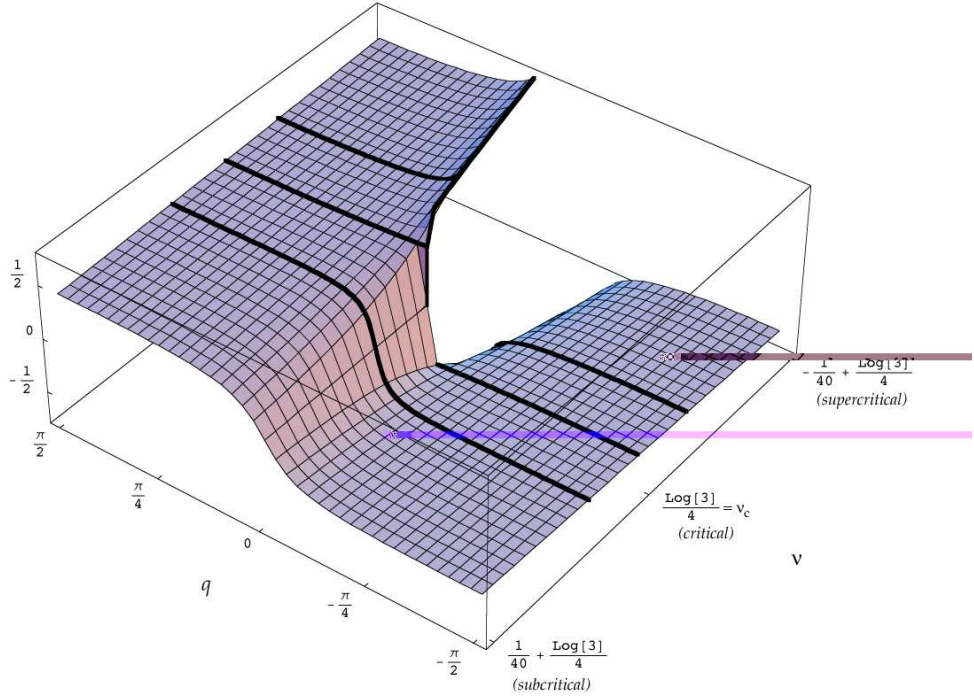}
\end{center}
\caption{The function ${\text{\rm Re\ }}d_{\text{\rm even}}(q,\nu)$ for the
triangular lattice.\label{fig:dPairTri}}
\end{figure}

We were not able to bring
the expressions for a generic temperature $\nu$ into a form suitable for
publication. Like the function $d$ of the square lattice (eq.\ (\ref{eq:cot})),
they seem to have an interval of discontinuity in the $(\nu,q)$-plane extending
from $(0,0)$ to $(\nu_c,0)$. This
is shown on Figure \ref{fig:dPairTri} where the real part of $d^{\text{\rm tri}}_{\text{\rm even}}$
has been plotted. 
Because of the singularity along the interval with $q=0$ and
$\nu\in(0,\nu_c]$, we have to limit the range (vertical axis) to the interval
$[-0.8,0.8]$. The cut on the surface imposed by this limitation
has been marked by a darker curve.
The restrictions of ${\text{\rm Re\ }}d^{\text{\rm tri}}_{\text{\rm even}}(\nu,q)$ to three
values of $\nu$ have been also highlighted. These three values are
$\nu_c+\frac1{40}=0.299653\dots$ in the subcritical phase,
the critical value $\nu_c=0.274653\dots$ and
$\nu_c-\frac1{40}=0.249653\dots$ in the supercritical phase. The subcritical
curve is obviously smooth. Lemma \ref{prop:lemma9} has established that the cricital
curve is real-analytic except at $q=0$ where there is a jump: ${\text{\rm Re\ }}
d^{\text{\rm tri}}_{\text{\rm even}}(\nu_c,0^+)=-{\text{\rm Re\ }}
d^{\text{\rm tri}}_{\text{\rm even}}(\nu_c,0^-)=2(2-\sqrt3)$.
The supercritical curve seems to have a pole at $q=0$.

%
%

\section*{Appendix: A combinatorial lemma}\label{sect:app}

The purpose of this appendix is to show the following lemma.

\begin{Lem}[Ising kernel] Let $m$ be a positive even integer and $n$ a positive integer
such that $2n\le m$. The polynomial $N_n(q_1, q_2,
\dots, q_n)$ in 
$e^{\pm iq_1},e^{\pm iq_2}, \dots, e^{\pm iq_n}$ defined by
\begin{equation}
\left(\prod_{1\le i\le n}{\text{\rm Odd}}_{q_i}\right){\text{\rm 
Sym}}_{q_1,q_2,
\dots,q_n}\sum_{1\le k_1<k_2<\dots<k_{2n}\le m}\ \sum_{\pi\in
\text{\rm Pf}_{2n}} (-1)^{l(\pi)}\prod_{1\le j\le n}e^{i(k_{\pi(2j-1)}-
k_{\pi(2j)})q_j}
\end{equation}
is 
$$=\begin{cases}
0,& \text{if there exists $i\neq j$ such that $q_i^2=q_j^2$},\\
{\displaystyle{\frac1{n!}\left(-\frac{im}2\right)^n\prod_{1\le j\le n}\cot
{\textstyle{\frac12}} q_j}},&
\text{otherwise},
\end{cases}
$$
when the values of the $q$'s are restricted to the set $Q_m=\{(2j-1)\pi/m, -\frac
m2+1\le j\le \frac m2\}$.
\end{Lem}

The notation is that of Section \ref{sect:flips}.

\noindent{\scshape Proof: } We first establish a one-to-one correspondence
between summands in \newline
$$\left(\prod_{1\le i\le n}{\text{\rm Odd}}_{q_i}\right){\text{\rm 
Sym}}_{q_1,q_2,
\dots,q_n}\sum_{\pi\in\text{\rm Pf}_{2n}}(-1)^{l(\pi)}
P(q_{\pi(1)},q_{\pi(2)},
\dots, q_{\pi(n)})$$ and in
$$\sum_{\rho\in S_{2n}}(-1)^{l(\rho)} P(q_{\rho(1)},q_{\rho(2)},\dots,
q_{\rho(n)}).$$
Take any permutation $\rho$ in $S_{2n}$. There exists a unique choice of $n$
permutations
$\tau_i\in S_{2n}$, acting trivially on all indices but $2i-1$ and $2i$ such
that 
$\tau_1(\rho(1))<\tau_1(\rho(2)), \quad
\tau_2(\rho(3))<\tau_2(\rho(4)),\quad \dots,\quad 
\tau_n(\rho(2n-1))<\tau_n(\rho(2n))$.
After the pairs have been ordered, there is a single permutation $\sigma\in
S_{2n}$ that orders pairs in increasing value of their first elements:
$\sigma(\tau_1(\rho(1)))< 
\sigma(\tau_2(\rho(3)))<\dots<
\sigma(\tau_n(\rho(2n-1)))$.
Because the choice of $\tau_i$'s amounts to choose one of the two
terms in $\text{\rm Odd}_{q_i}$ and the choice of $\sigma$
is one of the terms of the sum $\text{Sym}_{q_1,q_2,
\dots,q_n}$, the one-to-one correspondence 
stated above follows
with $\sigma \tau_1\tau_2\dots\tau_n\rho=\pi$. In
$\text{Odd}_{q_i}$, the term in which $q$ is replaced by $-q$ is
multiplied by $-1$ and the corresponding term in $\sum_{\rho\in S_{2n}}$
should appear with a factor $\prod_i (-1)^{l(\tau_i)}$. There is no
alternating sign in the operator $\text{\rm Sym}$. However, because $\sigma$
permutes pairs, it is an even permutation. Therefore
$(-1)^{l(\pi)}\prod_i (-1)^{l(\tau_i)}=(-1)^{l(\rho)}$
and 
\begin{equation}
N_n(q_1, q_2, \dots, q_n)=\frac1{2^n n!}\sum_{1\le k_1<k_2<\dots<k_{2n}\le m}\ 
\sum_{\rho\in S_{2n}}(-1)^{l(\rho)}\prod_{1\le j\le
n}e^{i(k_{\rho(2j-1)}-k_{\rho(2j)})q_j}.
\end{equation}
Note that by the introduction of auxiliary variables $x_{2j-1}=e^{iq_j}$ and $
x_{2j}=e^{-iq_j}$ for $1\le j\le n$, the sum over the permutation group $S_{2n}$ becomes simply
a determinant:
\begin{equation}
N_n(q_1, q_2, \dots, q_n)=\frac1{2^n n!}\sum_{1\le k_1<k_2<\dots<k_{2n}\le m}\ 
\det\left(x_i^{k_j}\right)_{1\le i,j\le 2n}.
\end{equation}

The second step is to transform the sum over ordered $2n$-tuplets $(k_1,k_2,
\dots,k_{2n})$ into a sum over partitions and to introduce the
appropriate tools from the theory of symmetric functions. 
A partition $\lambda$ is a finite set $\lambda=(\lambda_1,\lambda_2, \dots)$
of positive non-increasing integers.
(See \cite{Mac} for standard notations and definitions on
partitions.)  Define the integers
$\lambda_i, 1\le i\le 2n$ by
$\lambda_i=k_{2n-i+1}-(2n-i+1)$
or 
$k_j=j+\lambda_{2n+1-j}$.
The ordering $1\le k_1<k_2<\dots<k_{2n}\le m$ is equivalent to
$
m-2n\ge \lambda_1\ge\lambda_2\ge \dots\lambda_{2n}\ge 0$.
The partial ordering $\mu\subset\lambda$ between partitions stands for $\mu_i\le \lambda_i$, for all $i$.
Finally $(M^N)$ stands for the partition
$$(M^N)=(\underbrace{M,M,\dots,M}_{\text{\rm $N$ times}},0,0,\dots).$$
With this notation, the polynomial $N_n$ can be rewritten as
\begin{equation}
N_n(q_1, q_2, \dots, q_n)=\frac1{2^n n!}\ \sum_{\lambda\subset((m-2n)^{2n})}\ 
\det\left(x_i^{j+\lambda_{2n+1-j}}\right)_{1\le i,j\le 2n}.
\end{equation}
One can change the order of rows $(1,2,\dots, 2n)$ to $(2n, 2n-1, \dots, 1)$
in the determinant. The number of transpositions has the parity of 
$(2n)(2n-1)/2=n(2n-1)$ and has therefore the parity of $n$. Then
$
\det(x_i^{j+\lambda_{2n+1-j}})_{1\le i,j\le
2n}=(-1)^n\det(x_i^{\lambda_j+2n-j})_{1\le i,j\le 2n}\prod_{1\le j\le 2n}x_j$.
In the present case, the value of the $x_j$'s are such that
$x_{2j-1}=1/x_{2j}$ and the last product is simply unity. The definition
of the Schur function of $N$ variables associated with the partition $\lambda$ 
(also to be found in \cite{Mac}) is
$s_\lambda(x_1, x_2, \dots, x_N)=
{a_{\lambda+\delta}(x_1, x_2, \dots,x_N)}/{a_\delta(x_1, x_2, \dots, x_N)}$
with
$a_{\lambda+\delta}(x_1, x_2, \dots,x_N)=\det(x_i^{\lambda_j+N-j})_{
1\le i,j\le N}$
and $a_\delta$ is the Vandermonde determinant
$a_\delta(x_1, x_2, \dots, x_N)=\det(x_i^{N-j})_{
1\le i,j\le N}=\prod_{1\le i<j\le N}(x_i-x_j)$.
The polynomial $N_n$ is then
\begin{equation}
N_n(q_1, q_2, \dots, q_n)=\frac{(-1)^n}{2^n n!}\ 
\prod_{1\le i<j\le N}(x_i-x_j)
\sum_{\lambda\subset((m-2n)^{2n})} s_\lambda(x_1,x_2,\dots,x_{2n}).
\end{equation}
This form has the remarkable advantage over previous ones that the above sum
over partitions is known to combinatorists. On page 84 of \cite{Mac} one finds
$\sum_{\lambda\subset(M^N)}s_\lambda(x_1,x_2,\dots,x_N)={D_M}/{D_0}$
where
$D_M=\det(x_j^{M+2N-i}-x_j^{i-1})_{1\le i,j\le N}$
and $D_0=\det(x_j^{2N-i}-x_j^{i-1})_{1\le i,j\le N}$.
The computation of $N_n$ is then simply that of two determinants.

The third and last step is the computation of $D_0$ and $D_M$. Both shares
some obvious factors. Note that $M+2N=m+2n$ and $N=2n$ and their forms are
$$D_0=\left|
\begin{matrix}
x_1^{4n-1}-1   &   x_2^{4n-1}-1   &   \cdots   &   x_{2n}^{4n-1}-1      \\
x_1^{4n-2}-x_1 &   x_2^{4n-2}-x_2 &   \cdots   &   x_{2n}^{4n-2}-x_{2n} \\
\vdots           &   \vdots           &   \ddots   &   \vdots                 \\
x_1^{2n}-x_1^{2n-1} &   x_2^{2n}-x_2^{2n-1} &   \cdots   &   x_{2n}^{2n}-x_{2n}^{2n-1} 
\end{matrix}
\right|$$
and
$$D_{M}=\left|
\begin{matrix}
x_1^{m+2n-1}-1   &   x_2^{m+2n-1}-1   &   \cdots   &   x_{2n}^{m+2n-1}-1      \\
x_1^{m+2n-2}-x_1 &   x_2^{m+2n-2}-x_2 &   \cdots   &   x_{2n}^{m+2n-2}-x_{2n} \\
\vdots           &   \vdots           &   \ddots   &   \vdots                 \\
x_1^m-x_1^{2n-1} &   x_2^m-x_2^{2n-1} &   \cdots   &   x_{2n}^m-x_{2n}^{2n-1} 
\end{matrix}
\right|.$$
Each element of the $j$-th row of both $D_M$ and $D_0$
has an obvious $(x_j-1)$ factor. (Recall that $m\ge 2n$
and that $m+2n\ge 2i$ for $1\le i\le 2n$.) Both determinants
must contain a factor $\prod_{1\le j\le 2n} (x_j-1)$.
If $x_i=x_j$ for $i\neq j$, these determinants vanish. They 
must also contain a factor $\prod_{1\le i<j\le 2n}(x_i-x_j)$.
Finally, if one does $x_j=1/x_i$ in the $j$-th column of $D_M$, it becomes
\begin{align}
\left.(\text{$j$-th column})\right|_{x_j=1/x_i} &= (x_i^{-m-2n+k}-x_i^{-k+1})_{1\le k\le 2n}\\
&= -x_i^{-m-2n+1}(x_i^{m+2n-k}-x_i^{k-1})_{1\le k\le 2n}\\
&= -x_i^{-m-2n+1}(\text{$i$-th column})
\end{align}
and $D_M$ must contain also a factor $\prod_{1\le i<j\le 2n}(x_i x_j-1)$.
A similar computation for $D_0$ shows that it shares also this factor.

Let us define the {\em leading} monomial in a polynomial $P\in {\mathbb Q}[x_1,x_2,\dots, x_{2n}]$
as the one obtained through the following procedure. Among all monomials of $P$ find those that have
the largest exponent of $x_1$. Among the latter find those that have the largest exponent of $x_2$.
And so on till one chooses the one with the largest exponent of $x_{2n}$. This monomial is the
{\em leading} one. The leading monomial in $D_0$ is easy to determine. From the above determinantal
form, the largest exponent of $x_1$ is $4n-1$. Deleting the first column and row, we conclude
that the largest exponent of $x_2$ among the monomials containing $x_1^{4n-1}$ is $4n-2$.  And so
on. The leading monomial of $D_0$ is $\prod_{1\le j\le 2n}x_j^{4n-j}$ and its coefficient is 
unity. The leading monomial in the factors identified above in $D_0$
$$d_0=\prod_{1\le j\le 2n}(x_j-1)\prod_{1\le i<j\le 2n}(x_i-x_j)(x_i x_j-1)$$
can be also determined. The first product contributes $1$ to the exponent of $x_1$ in the leading
monomial and the second $2(2n-1)$. This exponent is therefore 
$(4n-1)$. Among those monomials containing
$x_1^{4n-1}$, the first product contributes $1$ to the exponent of $x_2$ and the second $(2n-2)+(2n-1)$
and the exponent of $x_2$ in the leading monomial is $4n-2$. Repeating the argument, it is
clear that the leading monomial in the factors above coincide with that of $D_0$ and its
coefficient is also $1$. We have thus shown that
$D_0=d_0$ and verified that $D_M$ has $D_0$ as factor (but this is a basic fact in the 
theory of Schur functions). Thus $d_m=D_M/D_0$ is a polynomial $\in {\mathbb Q}[x_1, x_2, \dots, 
x_{2n}]$. The polynomial $N_n$ has still a new form
$
N_n(q_1,q_2,\dots, q_n)=\frac{(-1)^n}{2^nn!}\, a_\delta\, d_m$.
This simple form has as immediate consequence that $N_n$ vanishes whenever $x_i=x_j$ with $i\neq j$, 
that is when there exists a pair $1\le k,l\le n$ of distinct integers such that $q_k=q_l$ ($i$ and
$j$ have the same parity) or $q_k=-q_l$ ($i$ and $j$ have different parity).

To calculate the value of $N_n$ when all $q_i^2$ are distinct, let us first remark that some of 
the factors in $a_\delta$ and in $D_0$ give easily the factors $\cot \frac12 q_j$ of the proposed
answer. Indeed the quotient of the factors $\prod_{1\le j\le n}(x_{2j-1}-x_{2j})$ in $a_\delta$
with the factors $\prod_{1\le j\le 2n}(x_j-1)$ of $D_0$ is
$$\frac{\prod_{1\le j\le n}(e^{iq_j}-e^{-iq_j})}{\prod_{1\le j\le n}(e^{iq_j}-1)(e^{-iq_j}-1)}=
\left(-\frac{2i}{(2i)^2}\right)^n \prod_{1\le j\le n}\frac{\sin q_j}{\sin^2 \frac12 q_j}=
i^n \prod_{1\le j\le n} \cot \frac12 q_j.$$
Let us denote by $\langle i,j\rangle$ pairs of odd integers with $1\le i<j<2n$. Then the residual factors
in $a_\delta$, that is those not used in the above quotient, are
$\prod_{\langle i,j\rangle} (x_i-x_j)(x_{i+1}-x_j)(x_i-x_{j+1})(x_{i+1}-x_{j+1})$
and those in $D_0$ are
$\prod_{1\le i<j\le 2n}(x_i-x_j)(x_i x_j-1)$
and
\begin{align}\label{eq:residu}
&\frac1{i^n\prod_{1\le j\le n} \cot \frac12 q_j}\frac{a_\delta}{D_0}\\
&\ =\frac1{\prod_{1\le j\le n}(x_{2j-1}-x_{2j})(x_{2j-1}x_{2j}-1)\notag
\prod_{\langle i,j\rangle} (x_ix_j-1)(x_{i+1}x_j-1)(x_ix_{j+1}-1)(x_{i+1}x_{j+1}-1)}.
\end{align}
The strategy will be to first simplify the factors $(x_{2j-1}-x_{2j})(x_{2j-1}x_{2j}-1)$ in
$D_M$ and then use freely the relation $x_{2j-1}=x_{2j}^{-1}$ in the remaining expression to
identify the last factors of \eqref{eq:residu}.

The factors $(x_{2j-1}-x_{2j})(x_{2j-1}x_{2j}-1)$ occurs in any $2\times2$ determinant of the
elements in the rows $(2k-1)$ and $(2k)$ and in the lines $i$ and $j$, $i<j$. If we set $x=
x_{2k-1}$ and $y=x_{2k}$, this is
$$d_{ij}(x,y)=\left|
\begin{matrix}
x^{m+2n-i}-x^{i-1}  &  y^{m+2n-i}-y^{i-1} \\
x^{m+2n-j}-x^{j-1}  &  y^{m+2n-j}-y^{j-1} 
\end{matrix}
\right|.$$
A direct calculation gives 
\begin{align*}
(x-y)(x y-1)\left( x^i y^{j-1}\sum_{k=0}^{j-i-1}\right. &(x/y)^k\sum_{l=0}^{m+2n-2-i-j}(xy)^l\\
& -\left.  x^{m+2n-2-j}y^i\sum_{k=0}^{m+2n-2-i-j}(y/x)^k\sum_{l=0}^{j-i-1}(xy)^l\right).
\end{align*}
After simplification with the factors $(x-y)(xy-1)=(x_{2j-1}-x_{2j})(x_{2j-1}x_{2j}-1)$ of
\eqref{eq:residu}, we can then use the identity $x=y^{-1}$ in the remaining expression
$d_{ij}(x,y)/(x-y)(xy-1)=\omega_{ij}(x)/(x-x^{-1})$
where
$$\omega_{ij}(x)=(m+2n-1-i-j)(x^{j-i}-x^{i-j})- (j-i)(x^{m+2n-1-i-j}-x^{-(m+2n-1-i-j)}).$$
Define the variables $y_j=x_{2j-1}$ (and $y_j^{-1}=x_{2j}$) for $1\le j\le n$. The remaining
factors of \eqref{eq:residu} are
\begin{align}
\prod_{\langle i,j\rangle} (x_ix_j-1)&(x_{i+1}x_j-1)(x_ix_{j+1}-1)(x_{i+1}x_{j+1}-1)\notag\\
&=
\prod_{1\le i<j\le n}(y_i y_j-1)(y_i/y_j-1)(y_j/y_i-1)(1/(y_iy_j)-1)\label{eq:starstar}
\end{align}
and the determinant $D_M$ has the following form in terms of the $\omega_{ij}$:
$$D_M=\frac1{\prod_{1\le j\le n}(y_j-y_j^{-1})}(n!\, \text{Sym}_{y_1, y_2, \dots, y_n})
\sum_{\pi\in \text{Pf}_{2n}} (-1)^{l(\pi)} \prod_{1\le j\le n}\omega_{\pi(2j-1),\pi(2j)}(y_i).$$
(Note that there are $(2n-1)!!$ permutations in $\text{Pf}_{2n}$ and that the symmetrization
operator gives rise to $n!$  terms out of each summand. The right-hand side contains
$(2n-1)!!n!=(2n)!/2^n$ terms. This is the right number as each of the $n$ factors $\omega$'s
is itself the sum of two terms.) We are now interested in the value of $D_M$ at $y_j$'s
such that none of the factors in the denominator is zero. We can therefore evaluate $D_M$
(and the $\omega$'s) at $y_j\in Q_m$, that is we can use in $D_M$ the fact that $y_j^m=-1$.
(This is the first time since $N_n$ has been expressed in terms of the quotient $D_M/D_0$
that this is possible.) The $\omega$'s are
$$\omega_{ij}(x)=(m+2n-1-i-j)(x^{j-i}-x^{i-j})+(j-i)(x^{2n-1-i-j}-x^{-(2n-1-i-j)})$$
for $x$ a $m$-root of $-1$. We can determine the leading monomial of the polynomial
$D_M$ (reduced by replacing all $y_j^m$ by $-1$) for the
variables $y_1, y_2, \dots, y_n$ taken in that order,
$y_1$ being the first.
Because $1\le i<j\le 2n$, the first term of any $\omega_{ij}$ reaches its largest value
$(2n-1)$ when $i=1$ and $j=2n$ and the other terms have smaller exponents than this maximum.
The leading term in $\omega_{1,2n}(y_1)$ is then $my_1^{2n-1}$. The range of $i$ and $j$
for $\omega_{ij}(y_2)$ is restricted to $2\le i<j\le 2n-1$. An analysis similar to the
one just done reveals that $i=2, j=2n-1$ are the values to choose and the corresponding
$\omega$ has leading term $my_2^{2n-3}$. This process leads to the leading monomial
in the above sum over $\pi\in\text{Pf}_{2n}$. It is $\prod_j y_j^{2n-(2j-1)}$. Its coefficient
is $m^n$. The corresponding monomial for $D_M$ is therefore $\prod_j y_j^{2(n-j)}$ with
the same coefficient.  (Note that the parity of the permutation putting
$(1,2n,2,2n-1,\dots, n-1,n)$ in increasing order is even and that it is Pfaffian.) 

Similarly, for the largest exponent of $y_1$ in \eqref{eq:starstar}, the first parenthesis contributes
$(n-1)$, the second also $(n-1)$ and the two last ones contribute nothing to the exponent
but gives a factor $(-1)^{2n-2}$ to its coefficient. The exponent is then
$(2n-2)$. Repeating the argument for the other variables, we find that
the leading monomial is the same as for $D_M$ and its coefficient is unity. We must
therefore conclude that, upon evaluation of the $\omega$'s at $m$-roots of
$-1$ whose squares are distinct, the quotient of $D_M$ with the residual factors \eqref{eq:residu} is independent of
the $q_j$ chosen and is $m^n$. The polynomial $N_n$ takes then the value
$\frac1{n!}(-\frac{im}2)^n\prod_{1\le j\le n} \cot
\frac12q_j$
whenever the squares $q_j^2$ are distinct. \hfill $\Box$

\bigskip

One has to note that the last calculation that rests upon the use of $y_j\in Q_m$ leads to
the wrong answer whenever $y_i=y_j$ or $y_i=y_j^{-1}$ for a given pair of distinct integers
$i$ and $j$. (The value of $N_n$ at these points was obtained earlier in the proof.)
This wrong result is not surprising as both $D_M$ and \eqref{eq:starstar} vanish at these points.
One has to cancel out common factors before using the property $y_j\in Q_m$. We have done
this exercise only for $m=4, n=2$ which is the first non-trivial case. In addition
to the $m^n=4^2$ terms computed above, one gets the following polynomial, 
after simplification of the common factors and {\em then} use of  $y_j\in Q_m$:
$-8-4xy+4x^3y+4x y^3-x^3y^3$ which, for $x,y\in Q_m$, is $-m^n (\delta_{xy,1}+\delta_{x/y,1})$.
It is obviously the right correction for the theorem to hold.

%
%

\section*{Acknowledments} 
 
The authors would like to thank Douglas Abraham, John Cardy, Anatole Joffe, Robert Langlands, Donald Richards, Thomas Spencer and Doron Zeilberger for
very helpful discussion. Y.\ S.-A.\ would also like to thank
the Institute for Advanced Study where part of this part was
done for its generosity and hospitality. L.-P. \ A. is supported by NSF grant DMS-0604869. L.-P.\ A. and Y.\ S.-A.\ acknowledge support from NSERC (Canada) and FCAR (Qu\'ebec).

%
%

\end{document}